\documentclass[structabstract]{aa}  
\usepackage{graphicx}
\usepackage{graphics}
\usepackage{txfonts}
\usepackage{natbib}
\begin{document}
\title{The abundance of HCN in circumstellar envelopes of AGB stars of different chemical type\thanks{This publication is based on data acquired with the Atacama Pathfinder Experiment (APEX) telescope, the IRAM 30\,m telescope, the James Clerk Maxwell Telescope (JCMT), the Swedish-ESO Submillimeter Telescope (SEST), and the Onsala 20\,m telescope. APEX is a collaboration between the Max-Planck-Institut fur Radioastronomie, the European Southern Observatory (ESO), and the Swedish National Facility for Radio Astronomy, Onsala Space Observatory (OSO). IRAM is supported by INSU/CNRS (France), MPG (Germany) and IGN (Spain). The JCMT is operated by the Joint Astronomy Centre on behalf of the Science and Technology Facilities Council of the United Kingdom, the Netherlands Organisation for Scientific Research, and the National Research Council of Canada. The Onsala 20\,m telescope is operated by OSO. The SEST was operated jointly by ESO and OSO.}}

\author{F. L. Sch\"oier\inst{1} \and S. Ramstedt\inst{2} \and H. Olofsson\inst{1} \and M. Lindqvist\inst{1} \and J. H. Bieging\inst{3} \and K. B. Marvel\inst{4}}

   \institute{Onsala Space Observatory, Dept. of Earth and Space Sciences, 
              Chalmers University of Technology, SE-43992 Onsala, Sweden
   \and Argelander Institut f\"ur Astronomie, University of Bonn, 53121 Bonn, Germany,
   \email{sofia@astro.uni-bonn.de}
   \and Steward Observatory, The University of Arizona, Tucson AZ 85721, USA
   \and American Astronomical Society, 2000 Florida Avenue NW, Suite 400, Washington, DC 20009, USA}    

   \date{Received ; accepted }


\abstract
{}{A multi-transition survey of HCN (sub-) millimeter line emission from a large sample of AGB stars of different chemical type is presented. The data are analysed and circumstellar HCN abundances are estimated. The sample stars span a large range of properties such as mass-loss rate and photospheric C/O-ratio. The analysis of the new data allows for more accurate estimates of the circumstellar HCN abundances and puts new constraints on chemical models. }
{In order to constrain the circumstellar HCN abundance distribution a detailed non-LTE excitation analysis, based on the Monte Carlo method, is performed. Effects of line overlaps and radiative excitation from dust grains are included.}
{The median values for the derived abundances of HCN (with respect to H$_{2}$) are 3\,$\times$\,10$^{-5}$, 7\,$\times$\,10$^{-7}$ and 10$^{-7}$ for carbon stars (25 stars), S-type AGB stars (19 stars) and M-type AGB stars (25 stars), respectively. The estimated sizes of the HCN envelopes are similar to those obtained in the case of SiO for the same sample of sources and agree well with previous results from interferometric observations, when these are available.}
{We find that there is a clear dependence of the derived circumstellar HCN abundance on the C/O-ratio of the star, in that carbon stars have about two orders of magnitude higher abundances than M-type AGB stars, on average. The derived HCN abundances of the S-type AGB stars have a larger spread and typically fall in between those of the two other types, however, slightly closer to the values for the M-type AGB stars. For the M-type stars, the estimated abundances are much higher than what would be expected if HCN is formed in thermal equilibrium. However, the results are also in contrast to predictions from recent non-LTE chemical models, where very little difference is expected in the HCN abundances between the various types of AGB stars.}

   
   \maketitle


\section{Introduction}
The final stages of the evolution of low- to intermediate-mass stars are, to a large extent, dominated by considerable mass loss in the form of an intense stellar wind. The circumstellar envelope (CSE), created by the wind as the star ascends the asymptotic giant branch (AGB), contains the products of the elements produced by internal nuclear processes, and dust grains formed in the inner part of the wind near the photosphere. Hence AGB mass loss contributes to the chemical evolution of the cosmic gas. In the outer CSE most molecules are photo-dissociated by the interstellar UV radiation field. The dust grains survive the radiation and there is the possibility that some of the more refractory molecules, such as e.g.\ SiO and SiS, are incorporated into grains and protected from photodissociation.

It is important to establish an understanding of the circumstellar molecular line emission from AGB stars and to derive accurate physical and chemical properties of the CSEs. The mass return of such stars, and hence their contribution to the galactic chemical evolution, is dominated by the rare high-mass-loss-rate objects. These are highly obscured at shorter wavelengths by their large amounts of circumstellar dust and their photospheric abundances can not be determined using classical methods, such as visual and near-IR spectroscopy. Instead, one must rely upon estimates based on circumstellar emission.

The structure of CSEs is relatively poorly known both on larger and on smaller scales. Single-dish observations do not generally resolve the sources and high resolution interferometric observations are limited at present.  Much of the current knowledge stems from interferometric observations of the nearby, high-mass-loss-rate carbon star IRC+10216 \citep[e.g.,][]{Bieging93, Dayal93, Dayal95, Lucas95, Fong06, Schoeier06c, Schoeier07b}, and the over-all picture of the morphology of AGB-CSEs is crude \citep{Neri98, castetal10}. The size of the emitting region is a crucial parameter for determining abundances, but theoretical estimates are uncertain, making interferometric observations important. Resolved images also test some of the basic assumptions of the radiative transfer model, e.g.,\ the validity of spherical symmetry and smoothness of the wind. 
 
Observed abundances and spatial distributions of the circumstellar molecular line emission will provide important tests of current chemical models of the stellar atmosphere and the CSE. The abundances of some common molecules such as HCN in carbon stars, and SiO in M-type AGB stars, can be reasonably well accounted for in chemical models where the molecules are formed in thermal equilibrium in, or near, the photosphere, and with little or no further processing until they are eventually photo-dissociated by the ambient UV-field. The detection of carbon-bearing species such as HCN, CN, CS and H$_2$CO  in envelopes of M-type AGB stars \citep[e.g, ][]{Deguchi85, Lindqvist88, Olofsson91,Lindqvist92}
and oxygen-bearing molecules like H$_2$O, H$_2$CO, C$_3$O and SiO in envelopes of carbon stars \citep[e.g.,][]{Keady93,Melnick01,Ford04,Hasegawa06,Schoeier06a, Tenenbaum06,neufetal11} points to the importance of non-equilibrium chemical processes. Models in which periodic shock waves propagate in the extended photosphere \citep{willcher98,duaretal99,Agundez06,Cherchneff06,Cherchneff11} appear to explain many of these observed abundances, however, a more statistical study is still lacking. Other molecular abundances may also be significantly affected by the formation of dust grains in the inner part of the circumstellar envelope (e.g., Ag\'undez \& Cernicharo 2006, SiO and H$_{2}$O; Sch\"oier et al.~2006b, SiO)\nocite{Agundez06,Schoeier06a}.

HCN has been detected in AGB CSEs of all chemical types \citep[e.g.,][]{Bieging00}, and its formation is sensitive to shocks passing through the gas, making it a good probe of the important chemical processes \citep{Cherchneff06}. \citet{Lindqvist00} observed a sample of four carbon stars in the HCN $J$\,=\,1\,$\rightarrow$\,0 and CN $N$\,=\,1\,$\rightarrow$\,0 lines, using the IRAM Plateau de Bure interferometer, in an attempt to constrain their abundance distributions. From detailed numerical modelling they found a qualitative agreement with current theoretical models of circumstellar chemistry when it comes to the size estimates (though, the statistics are poor and sources with intermediate mass-loss rates are missing). The spatial extents of the HCN and CN envelopes derived from the radiative transfer analysis of the observational data, are systematically larger than predictions from photodissociation modelling, by about a factor of two. Observations of circumstellar HCN line emission from the two M-type AGB stars TX Cam and IK Tau by \citet{Marvel05} clearly showed that HCN is also of photospheric origin in M-type AGB stars, which resolved a long-standing debate. \citet{Muller08} observed HCN $J$\,=\,8\,$\rightarrow$\,7 emission, using the SMA, towards the M-type AGB star W Hya. Their data corroborates the findings of \citet{Marvel05} as to the origin of the HCN emission, and also probes gas still accelerating in the inner wind of the star. The recent analysis of the HCN Herschel/HIFI data (in combination with groundbased data) of the S-star $\chi$~Cyg \citep{schoetal11} also suggest a more intricate HCN distribution.

In this paper we present new multi-transition observations of HCN line emission towards a sample of AGB stars with different properties, and circumstellar fractional HCN abundances (with respect to H$_{2}$, hereafter referred to as abundance) derived using a detailed radiative transfer model. Together with existing data, these observations will significantly extend the observed range of parameter space, and provide a solid statistical test of current chemical models in and near the photosphere of AGB stars, as well as further out in the circumstellar envelopes.


\section{Observations and data reduction}

\begin{table}
\caption{{\bf Telescope data relevant for the new observations of HCN.}}
\label{efficiencies}
$
\begin{array}{cccccccccccc}
\hline
\noalign{\smallskip}
\multicolumn{1}{c}{{\mathrm{Transition}}} & &
\multicolumn{1}{c}{{\mathrm{Frequency}}} &  &
\multicolumn{1}{c}{{E_{\mathrm{up}}}} &  &
\multicolumn{1}{c}{{\mathrm{Telescope}}}  &&
\multicolumn{1}{c}{\eta_{\mathrm{mb}}}  && 
\multicolumn{1}{c}{\theta_{\mathrm{mb}}} \\ 
& & 
\multicolumn{1}{c}{{\mathrm{[GHz]}}} && 
\multicolumn{1}{c}{{\mathrm{[K]}}} &&  
&& &&
\multicolumn{1}{c}{[\arcsec]} \\
\noalign{\smallskip}
\hline
\noalign{\smallskip}
J=1\rightarrow0 && \phantom{0}88.632 && \phantom{00}4 && \mathrm{IRAM} && 0.78 && 28 \\
 && && && \mathrm{NRAO} && 0.90 && 71 \\
  && && && \mathrm{OSO} && 0.60 && 43 \\
 && && && \mathrm{SEST} && 0.75 && 56 \\
J=3\rightarrow2 && 265.886 && \phantom{0}26 && \mathrm{APEX} && 0.75 && 23 \\  
 && && && \mathrm{IRAM} && 0.53 && 10 \\
  && && && \mathrm{JCMT} && 0.69 && 19 \\
 && && && \mathrm{NRAO} && 0.45 && 27 \\ 
 && && && \mathrm{SEST} && 0.50 && 20 \\
J=4\rightarrow3 && 354.505 && \phantom{0}43 && \mathrm{APEX} && 0.70 && 18 \\
 && && && \mathrm{JCMT} && 0.60 && 14 \\
 && && && \mathrm{SEST} && 0.25 && 15 \\
\noalign{\smallskip}
\hline
\end{array}
$
\end{table}


\label{single-dish}
{\bf New m}ulti-transition HCN line observations (see Table~\ref{efficiencies}) were performed during 1995--2008 using the Onsala 20\,m telescope (OSO), the Swedish-ESO 15\,m submillimetre telescope (SEST), the JCMT 15\,m telescope, the APEX 12\,m telescope, the IRAM 30\,m telescope, and the NRAO 12\,m telescope at Kitt Peak, USA.

The OSO, SEST, JCMT, and IRAM observations were made in a dual beamswitch mode, 
where the source is alternately placed in the signal and the reference beam, using a beam throw of about $11\arcmin$ (OSO and SEST) or $2\arcmin$ (JCMT and IRAM). This method produces very flat baselines. At APEX and NRAO, the observations were carried out using a position-switching mode, with the reference position located +3$\arcmin$ and +10$\arcmin$ in azimuth, respectively. Regular pointing checks were made on SiO masers (OSO and SEST) and strong CO and continuum sources (APEX, JCMT, IRAM, and NRAO). Typically the pointing was found to be consistent with the pointing model within $\approx$\,3$\arcsec$ for all telescopes, except at the NRAO where it was found to be consistent within $\approx$\,5$\arcsec$. All receivers were single sideband receivers.

The raw spectra are stored in $T_{\mathrm A}^{\star}$-scale and converted to main-beam brightness temperature using $T_{\mathrm{mb}}$\,=\,$T_{\mathrm A}^{*}/\eta_{\mathrm{mb}}$. $T_{\mathrm A}^{\star}$ is the antenna temperature corrected for atmospheric attenuation using the chopper-wheel method, and $\eta_{\mathrm{mb}}$ is the main-beam efficiency. The adopted beam efficiencies, together with the FWHM:a of the main beams ($\theta_{\mathrm{mb}}$), for all telescopes and frequencies are given in Table~\ref{efficiencies}. The uncertainty in the absolute intensity scale is estimated to be about $\pm 20$\%.  In Table~\ref{efficiencies}, the energy of the upper level involved in the transition ($E_{\mathrm{up}}$) is also given, ranging from 4\,K for the $J$\,$=$\,1 level up to 43\,K for the  $J$\,$=$\,4 level, illustrating the potential of these multi-transition observations to probe a relatively large radial range of the CSE (see Sect.~\ref{sect_size}).

The data were reduced by removing a low (typically first) order polynomial baseline and then binned (typically to a velocity resolution of about 1\,km\,s$^{-1}$) in order to improve the 
signal-to-noise ratio, using XS\footnote{XS is a package developed by P. Bergman to reduce and analyse single-dish spectra. It is publicly available from {\tt ftp://yggdrasil.oso.chalmers.se}}. The observed spectra are presented in Figs~\ref{spectrac1}--\ref{spectram} and velocity-integrated intensities are reported in Tables~\ref{intensities1} \& \ref{intensities2}. The intensities are given in main-beam brightness temperature scale ($T_{\mathrm{mb}}$). In some of the carbon-star spectra, the $\nu_2$\,=\,1, $J$\,=\,3\,$\rightarrow$\,2, and $J$\,=\,4\,$\rightarrow$\,3 lines can be seen at slightly lower frequency than the corresponding ground-state lines (especially apparent in the spectra of e.g., Y~CVn and CW~Leo). The same vibrationally excited line appears to be present in the $J$\,=\,3\,$\rightarrow$\,2 spectrum of the M-type star R~Leo, however, it does not appear exactly at the correct frequency, and when re-observed it was not detected. 

In addition to the new data presented here we have also used HCN line intensities reported by \citet{Olofsson93b}, \citet{Olofsson98b}, \citet{Bujarrabal94}, \citet{Bieging00}, and \citet{Woods03}, as well as data from the JCMT public archive\footnote{{\tt http://www.jach.hawaii.edu/JACpublic/JCMT/}}. All the data used in the excitation analysis are included in Tables~\ref{intensities1} \& \ref{intensities2}.


\section{Radiative transfer modelling}
\label{sect_model}
\subsection{General assumptions}
The CSEs are assumed to be spherically symmetric, produced by a constant mass-loss rate ($\dot{M}$), and expanding at a constant velocity ($\upsilon_{\mathrm e}$). The molecular line excitation analysis is performed using a detailed non-LTE radiative transfer code, based on the Monte Carlo method and described in detail in \citet{Schoeier01}. The code has been extensively tested and benchmarked against a wide variety of molecular line radiative transfer codes in \citet{Zadelhoff02}. The local line width is assumed to be described by a Gaussian and is made up of a micro-turbulent component with a Doppler width of 1.0\,km\,s$^{-1}$ and a thermal component that is directly calculated from the derived kinetic temperature structure (see Sect.~\ref{co_mod}).

The analysis includes radiative excitation through vibrationally excited states of CO and HCN. Both the central star (approximated by a blackbody), and thermal dust grains distributed in the CSE, can provide radiation fields capable of populating vibrationally excited states, and thereby affect the excitation in the ground vibrational state. The addition of a dust component in the Monte Carlo scheme is straightforward, as described in \citet{Schoeier02b}. 

The best-fit model is found by minimizing the total $\chi^2$ defined as
\begin{equation}
\label{chi2_sum}
\chi^2_{\mathrm{tot}} = \sum^N_{i=1} \left [ \frac{(I_{\mathrm{mod}}-I_{\mathrm{obs}})}{\sigma}\right ]^2, 
\end{equation} 
where $I$ is the velocity-integrated line intensity and $\sigma$ the uncertainty in the measured value. The $\sigma$-value is usually dominated by the calibration uncertainty assumed to be $\pm$20\%. In general, the data quality is good and an uncertainty of 20\%  is a reasonable estimate.  However, it is possible that the error is larger in some cases, and for very noisy lines we have added a higher uncertainty, thereby assigning a lower weight to the noisy lines in the modelling. The summation is done over $N$ independent observations. When modelling the dust continuum radiative transfer (Sect.~\ref{sed_mod}), Eq.~\ref{chi2_sum} is also used to find the best-fit model, but with the flux density, $F_{\lambda}$, instead of $I$.


\subsection{Dust continuum emission modelling}
\label{sed_mod}
The dust temperature structure and dust density profile are obtained from detailed radiative transfer modelling using the publicly available code {\em Dusty} \citep{Ivezic97}.  Here we are mainly interested in the potential effect that dust grains have on the excitation of molecules. A full treatment of dust radiative transfer coupled with a dynamical model for the wind, in order to independently measure physical properties of the CSE, such as e.g. the mass-loss rate, is beyond the scope of this paper [see e.g. \citet{schoetal11} for an example of such modelling]. In the modelling, where the spectral energy distribution (SED) provides the observational constraint, the dust optical depth specified at 10\,$\mu$m, $\tau_{10}$, the dust temperature at the inner radius of the dust envelope, $T_{\mathrm{d}} (r_{\mathrm{i}})$, and the effective stellar blackbody temperature, $T_{\star}$, are the adjustable parameters. The SED is typically constrained by {\em{JHKLM}} photometric data [from 2MASS, \citet{Kerschbaum99b}, and Kerschbaum priv. com.], IRAS fluxes, and in some cases sub-millimetre data \citep{Groenewegen93}. As mentioned above, the best-fit model is found by minimizing Eq.~\ref{chi2_sum}.

The total luminosity, $L_{\star}$, for regularly pulsating stars, is obtained from the period-luminosity relations of \citet[][for carbon stars]{Groenewegen96} and \citet[][for M- and S-type AGB stars]{Whitelock94}. In the case of non-regularly pulsating stars, a generic luminosity of 4000\,L$_{\sun}$ is adopted. The distance, $D$, is then obtained from the SED-fitting once the luminosity of the star is known. For the sources where reliable Hipparcos parallaxes exist, the corresponding distance has been adopted and the luminosity is then, in turn, obtained from the SED modelling. Amorphous carbon dust grains with the optical constants given in \citet{Suh00} are adopted for the carbon stars and astronomical silicates \citep{Justtanont92} for the M-type AGB stars. For the S-type AGB stars, either carbon or silicate grains are assumed based on the IRAS low-resolution spectra classification \citep{Volk89}. For simplicity, the dust grains are assumed to have the same radius, $a_{\mathrm{d}}$=0.1\,$\mu$m, and the same mass density, $\rho_{\mathrm{d}}$\,=\,2.0\,g\,cm$^{-3}$ (for carbon grains) and 3.0\,g\,cm$^{-3}$ (for silicate grains). The corresponding dust opacities, $\kappa_{\nu}$, are calculated from the optical constants and the individual grain properties using standard Mie theory \citep{Bohren83}. 

The results from the dust continuum emission modelling  are listed in Tables~\ref{sample_c} and \ref{sample_sm}. The results for the S-type AGB stars are taken from \citet{Ramstedt09}.


\subsection{CO line modelling}
\label{co_mod}
The physical properties of the circumstellar gas, such as the density, temperature, and kinematic structures, are derived from radiative transfer modelling of multi-transition millimetre and sub-millimetre CO line observations. The CO data used in the analysis have been presented in \citet{Schoeier01}, \citet{Olofsson02}, \citet{Delgado03b}, \citet{Schoeier06a} and \citet{Ramstedt06,Ramstedt08,Ramstedt09}. The abundance distribution of CO is based on the photochemical models of \citet{Mamon88} adopting an initial (photospheric) abundance of 1\,$\times$\,10$^{-3}$ for carbon stars, 6\,$\times$\,10$^{-4}$ for S-type AGB stars and 2\,$\times$\,10$^{-4}$ for M-type AGB stars. The kinetic temperature structure is obtained in a self-consistent manner from solving the energy balance equation, where the CO line cooling is directly obtained from the excitation analysis. Mass-loss rates derived from detailed radiative transfer modelling of CO multi-transition line emission are accurate to within a factor 2--4 \citep{Ramstedt08}.

The excitation analysis includes the first 41 rotational levels within the ground state and the first vibrationally excited ($v$\,=\,1) state of CO at 4.6\,$\mu$m. The collisional rates are taken from \citet{Flower01a} and \citet{Wernli06}. They were calculated for temperatures in the range from 5 to 400\,K including rotational energy levels up to $J$\,=\,29 and $J$\,=\,20 for collisions with para-H$_2$ and ortho-H$_2$, respectively. The collisional rate coefficients have been extrapolated to include energy levels up to $J$\,=\,40 and collisional temperatures up to 2000\,K, as described in \citet{Schoeier05a}, and are made publicly available through the {\em Leiden Atomic and Molecular Database} (LAMDA){\footnote{\tt http://www.strw.leidenuniv.nl/$\sim$moldata}}. An ortho-to-para ratio of 3 was adopted when weighting together collisional rate coefficients for CO in collisions with ortho-H$_2$ and para-H$_2$.

The parameters obtained from the CO excitation analysis are reported in Tables~\ref{sample_c} and \ref{sample_sm}.

 
\subsection{HCN line modelling}
Once the basic envelope parameters are known from the CO line modelling and dust radiative transfer, the abundance of HCN in the circumstellar envelopes can be determined. The HCN abundance distribution (i.e. the ratio of the number density of HCN molecules to that of H$_2$ molecules as a function of radius), $f$\,$=$\,$n\mathrm{(HCN)}/n\mathrm{(H_2)}$, is assumed to be described by a Gaussian
\begin{equation}
\label{eq_distr}
f(r) = f_0\, \exp \left(-\left(\frac{r}{r_{\mathrm e}}\right)^2 \right),
\end{equation}
where  $f_0$ is the initial abundance, and $r_{\mathrm e}$ the e-folding radius for HCN. 

HCN is a linear molecule and has three different vibrational modes: the CN stretching mode $\nu_1$, the CH stretching mode $\nu_3$, and the doubly degenerate bending mode $\nu_2$ (doubly degenerate since the bending of a non-rotating linear molecule is direction independent and will result in radiation of the same frequency). The excitation analysis includes radiative excitation through the CH stretching mode ($\nu_3$\,=\,1;  also denoted 001) at 3\,$\mu$m, and the bending mode ($\nu_2$\,=\,1;  also denoted 010) at 14\,$\mu$m. The CN stretching mode ($\nu_1$\,=\,1;  also denoted 100) at 4.8\,$\mu$m has transitions that are about 20\,--\,1000 times weaker and it is therefore not included in the analysis. In each of the vibrational levels we include rotational levels up to $J$\,=\,29.  

The first vibrational overtone (020) lies at energies above the ground state corresponding to 7\,$\mu$m and has transitions to the (010) and (000) states with Einstein A coefficients of the same order as those from the (010) to the (000) state \citep[e.g.,][]{ziurturn86}. A test was performed to estimate its influence on the excitation analysis covering optical depths that are characteristic for the sample sources. For all models, the total integrated intensities of the lines included in this paper are changed by less than a few percent when the (020) state is included. At high optical depths the radiative excitation is of minor significance and therefore likely also the excitation to higher vibrational states. An effect on higher rotational lines at intermediate and low optical depths, cannot be excluded, but to evaluate this is beyond the scope of this paper. For the excitation analysis and intensities of the lines used to constrain the modelling performed in this paper, we conclude that excitation through the (020) state will have no significant effect.

The nuclear spin of the $^{14}$N atom results in the splitting of the rotational levels into three hyperfine components. Hyperfine splitting is included only for rotational levels in the ground state up to $J$\,=\,4 (where the splitting is of the same order as the local line width) and the resulting line overlaps are accurately treated as described in Lindqvist et al.\ (2000). Frequencies for the hyperfine transitions are obtained from the CDMS database \citep{Muller01, Muller05}.

When the molecule is bending and rotating simultaneously, the double degeneracy of the bending mode will be lifted, resulting in the splitting of the bending mode (010) into two levels (01$^{1c}$0 and 01$^{1d}$0). This is referred to as $l$-type doubling and is included in the excitation analysis. The transition probabilities and frequencies are taken from the GEISA database \citep{Husson05}. 

The collisional rate coefficients are taken from \citet{Green74} and Green (unpublished data{\footnote{\tt http://data.giss.nasa.gov/mcrates/data/hcn\_he\_rates.txt}}). The rate coefficients presented in \citet{Green74} were calculated for temperatures in the range from 5 to 100\,K, including energy levels up to $J$\,=\,7 for collisions with He. The unpublished rates by Green were obtained for the lowest 30 rotational transitions and for temperatures from 100 to 1200\,K, again for collisions with He. These two sets of rate coefficients were subsequently scaled by a factor 1.37 to represent collisions with H$_2$. Extrapolation to include all energy levels up to $J$\,=\,29 for temperatures between 5 and 1200\,K was made and can be obtained from the LAMDA database \citep{Schoeier05a}. Quite recently, new collision rates were published by \citet{dumoetal10}. We tested using these rates for three objects with low, intermediate, and high mass-loss rate. The results differed by less than a few percent compared to the results when using the scaled and extrapolated data from \citet{Green74}, for all three cases.

The results from the HCN excitation analysis are given in Tables~\ref{sample_c} and \ref{sample_sm}.


\section{Results and discussion}
\label{res_and_dis}
\subsection{Results from the radiative transfer modelling}
In order to determine accurate abundances it is important to know the spatial extent of the molecular envelope. Since different molecular transitions have different excitation conditions, multi-transition observations have the potential to constrain both the initial abundance, $f_0$, and e-folding radius, $r_{\mathrm e}$ (Eq.~\ref{eq_distr}) and hence describe the distribution of HCN molecules in the CSE. The best-fit initial abundances provide constraints on chemical models in and near the photosphere as well as dust condensation models. The envelope sizes obtained, on the other hand, constrain photochemical models of the CSE. This is illustrated in \citet{Delgado03b}, \citet{Schoeier06a}, and \citet{Ramstedt09} for SiO, and in \citet{Schoeier07a} for SiS. 

Due to lack of observational constraints a full analysis, determining both the initial abundance and the envelope size, is possible only for a sub-sample (20) of sources. For these sources, both parameters are considered as free parameters and varied within reasonable ranges until the best fit to the observations are found. The results are then used to derive a relation describing how the HCN envelope size scales with the wind density ($\dot{M}/v_{\rm{e}}$). This relation is used to estimate the HCN envelope size for all sources and the initial abundance is the only free parameter in the radiative transfer analysis. Given the uncertainties involved in the modelling of carbon stars as explained further in Sect.~\ref{considerations} (particularly for the $J$\,=1\,$\rightarrow$\,0 transition) no carbon stars are used when deriving the HCN envelope size relation.


\subsubsection{HCN envelope size}
\label{sect_size}
\begin{figure*}[t]
      \centerline{\includegraphics[width=15cm]{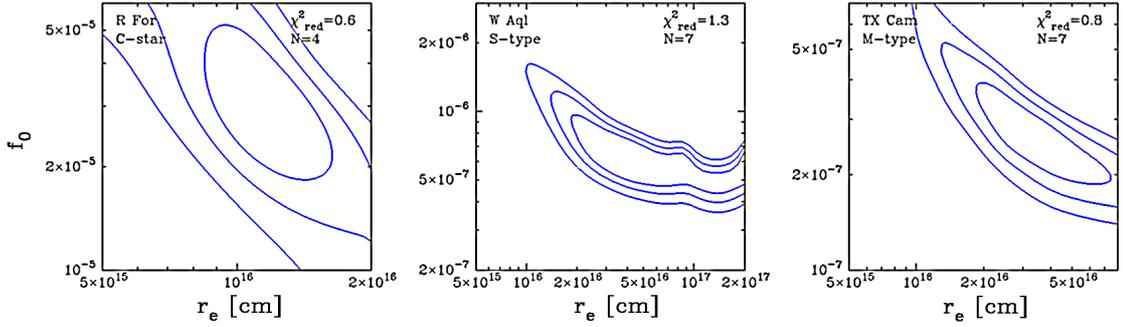}}
   \caption{$\chi^2$-map showing the quality of the fit to available HCN multi-transition single-dish data for the AGB stars R For (left panel), W Aql (middle panel), and TX Cam (right panel) when varying the adjustable parameters, $f_0$ and $r_{\mathrm{e}}$, in the model. Contours are drawn at the 1, 2, and 3\,$\sigma$ levels. Indicated in the upper right corner is the reduced $\chi^2$ of the best-fit model and the number of observational constraints used, $N$.}
      \label{txcam_chi2}
      \end{figure*}

\begin{figure*}[h]
      \centerline{\includegraphics[width=16cm]{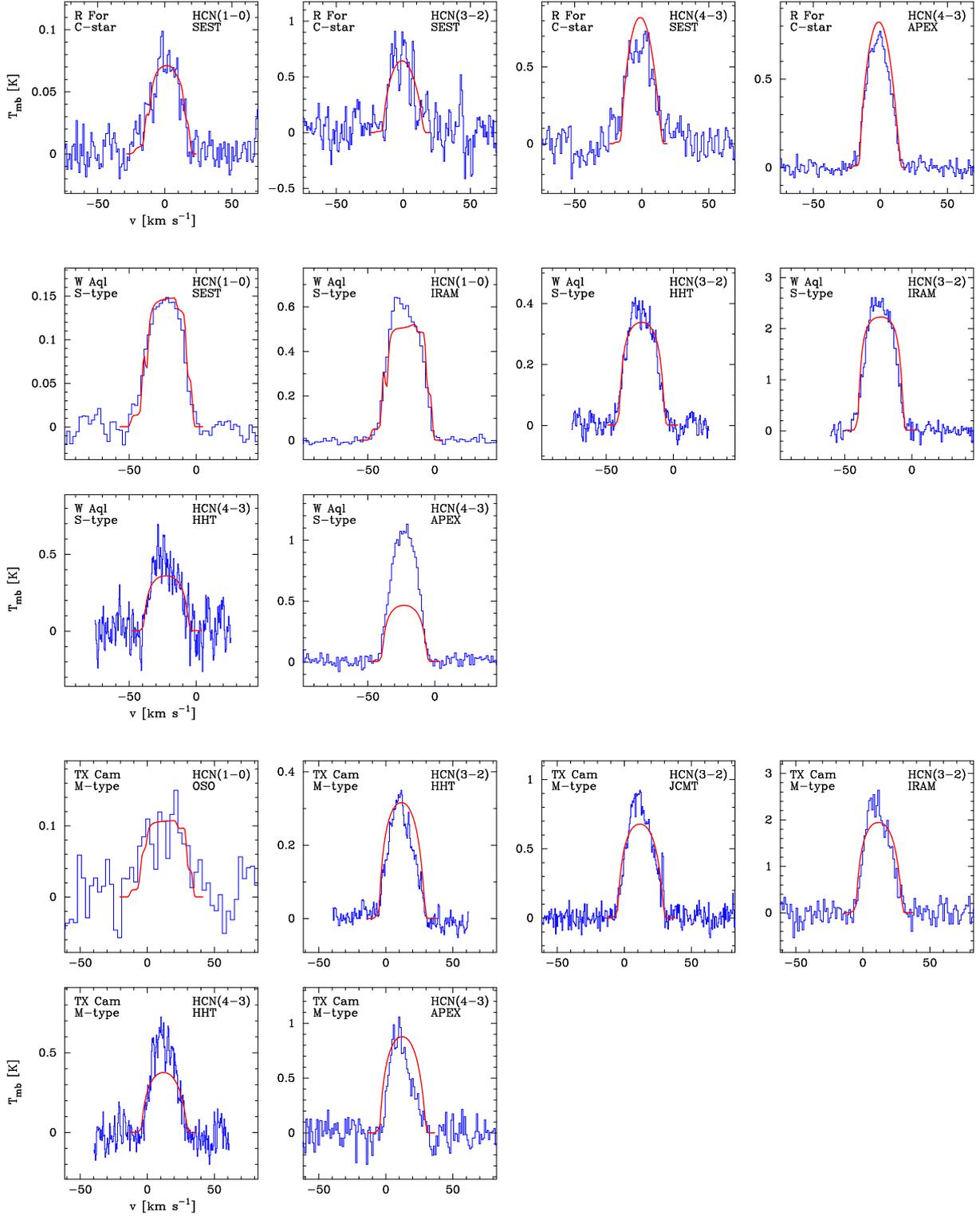}}
   \caption{Multi-transition spectra (histograms) of HCN line emission (HHT data from Bieging et al.\ 2000), overlaid by spectra from the best-fit model (solid lines; assuming a Guassian abundance distribution), towards {\bf top panels --} the carbon star R For using a fractional HCN abundance $f_0$\,$=$\,4.0\,$\times$\,10$^{-5}$ and an envelope size $r_{\mathrm{e}}$\,$=$\,9.3\,$\times$\,10$^{15}$\,cm. {\bf middle panels --} S-type AGB star W Aql using a fractional HCN abundance $f_0$\,$=$\,1.2\,$\times$\,10$^{-6}$ and an envelope size $r_{\mathrm{e}}$\,$=$\,1.3\,$\times$\,10$^{16}$\,cm.  {\bf bottom panels --} M-type AGB star TX Cam using a fractional HCN abundance $f_0$\,$=$\,3.5\,$\times$\,10$^{-7}$ and an envelope size $r_{\mathrm{e}}$\,$=$\,2.0\,$\times$\,10$^{16}$\,cm are also shown (solid lines). 
   The hyperfine splitting of the $J$\,=\,1\,$\rightarrow$\,0 transition significantly broadens the line. This effect is explicitly taken into account in the modelling. The calibration uncertainty in the observed spectra is $\approx$\,$\pm$\,20\%.}
      \label{txcam_spectra}
      \end{figure*}

In the present sample 12 M- and 8 S-type AGB stars have been observed in three or more different lines. For these sources, the size of the HCN radio-line-emitting region ($r_{\mathrm e}$) and the inner HCN abundance ($f_0$) are varied simultaneously. In 16 cases it is possible to constrain both parameters reasonably well, as reported in Table~\ref{table_chi2}. The sensitivity of the line emission to variations of the two adjustable parameters $f_0$ and $r_{\mathrm e}$ is illustrated in Fig.~\ref{txcam_chi2} for the high-mass-loss-rate M-type AGB star \object{TX~Cam} and the intermediate-mass-loss-rate S-type AGB star \object{W~Aql}. In Fig.~\ref{txcam_spectra}, the model spectra are directly compared with the observations for the best-fit models. The results for the intermediate-mass-loss-rate carbon star  \object{R~For} are also shown in Figs~\ref{txcam_chi2} and \ref{txcam_spectra} as an example, however they are not included when deriving the HCN envelope size relation. In most cases the reduced $\chi^2$ values ($\chi^2_{\mathrm{red}} = \chi^2_{\mathrm{tot}}/(N-2)$) are $\leq$\,3 indicating reasonably good fits to the observed HCN line emission. 

\begin{table}
\caption{HCN radial abundance distribution results.}
\label{table_chi2}
$$
\begin{array}{p{0.2\linewidth}cccc}
\hline
\noalign{\smallskip}
&
\multicolumn{1}{c}{{f_0} ^a} & 
\multicolumn{1}{c}{{r_{\mathrm{e}}} ^a} & 
\multicolumn{1}{c}{N}  &
\multicolumn{1}{c}{\chi^2_{\mathrm{red}}}  \\ 

\multicolumn{1}{l}{{\mathrm{Source}}} &
 &
\multicolumn{1}{c}{[{\mathrm{cm}}]} & &\\
\noalign{\smallskip}
\hline
\noalign{\smallskip}
\ {\em C-star} \\
\object{R For}            &  3.5\pm1.5\times10^{-5} &  1.2\pm0.4\times10^{16} & 4 & 0.6\\
\ {\em S-type} \\
\object{R And}            &  2.2\pm1.2\times10^{-7} &  4.8\pm4.0\times10^{16} & 6 & 4.0\\
\object{W And}            &  5.0\pm3.0\times10^{-7} &  4.3\pm2.2\times10^{15} & 3 & 5.5\\
\object{W Aql}             &  7.0\pm2.5\times10^{-7} &  5.8\pm4.0\times10^{16} & 7 & 1.3\\
\object{$\chi$ Cyg}    &  3.0\pm1.1\times10^{-6} &  5.5\pm2.5\times10^{15} & 4 & 0.9\\
\object{R Cyg}            &  2.7\pm1.0\times10^{-7} &  4.5\pm3.5\times10^{16} & 4 & 6.1\\
\object{S Lyr}              &  4.5\pm3.4\times10^{-6} &  2.2\pm1.7\times10^{16} & 3 & 4.5\\
\ {\em M-type} \\
\object{RX Boo}          &  4.7\pm2.3\times10^{-7} &  3.5\pm1.4\times10^{15} & 6 & 2.2\\
\object{TX Cam}         &  2.8\pm1.1\times10^{-7} &  4.5\pm2.7\times10^{16} & 7 & 0.8\\
\object{R Cas}            &  3.5\pm1.4\times10^{-7} &  5.9\pm2.1\times10^{15} & 6 & 3.5\\
\object{R Dor}             &  9.0\pm4.0\times10^{-8} &  7.1\pm4.0\times10^{15} & 3 & 2.0\\
\object{W Hya}            &  8.5\pm4.5\times10^{-7} &  3.0\pm1.1\times10^{15} & 4 & 6.1\\
\object{R Leo}             &  2.5\pm1.1\times10^{-7} &  6.0\pm2.5\times10^{15} & 4 & 0.9\\
\object{WX Psc}          &  3.0\pm2.0\times10^{-7} &  3.0\pm2.0\times10^{16} & 5 & 1.0\\
\object{IK Tau}            &  4.3\pm2.8\times10^{-7} &  1.4\pm0.8\times10^{16} & 4 & 0.5\\
\object{IRC-10529}    &  4.5\pm2.5\times10^{-8} &  3.2\pm1.8\times10^{16} & 4 & 2.7\\
\object{IRC-30398}    &  1.1\pm0.3\times10^{-7} &  2.6\pm1.3\times10^{16} & 4 & 0.9\\
\noalign{\smallskip}
\hline
\end{array}
$$
$^a$ The abundance distribution is assumed to be Gaussian (see Eq.~\ref{eq_distr})
\end{table}

   \begin{figure}
	\centerline{\includegraphics[width=7cm]{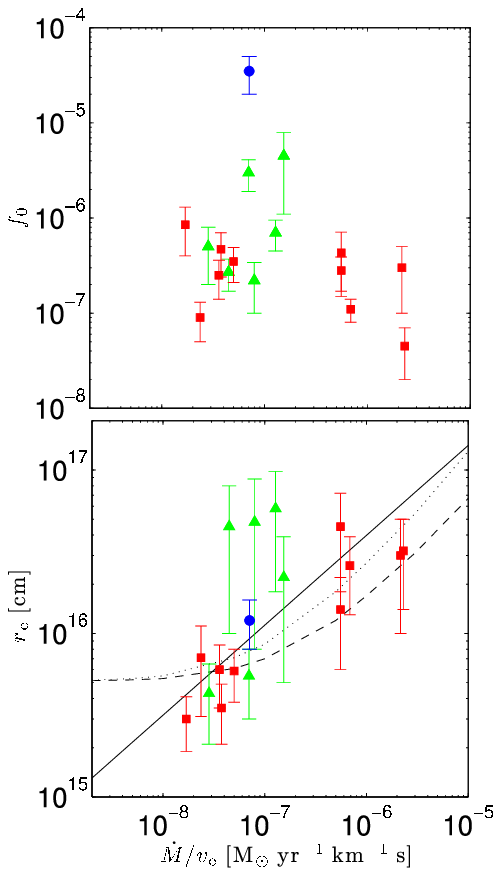}}
   \caption{Results from excitation analysis of HCN multi-transition single-dish observations assuming a Gaussian abundance distribution (see Eq.~\ref{eq_distr}). {\em Top panel --} Derived initial fractional abundance ($f_0$) as a function of envelope density ($\dot{M}/v_{\mathrm{e}}$) for M- (squares) and S-type AGB stars (triangles). {\em Bottom panel --} Derived envelope sizes ($r_{\mathrm{e}}$) as a function of envelope density ($\dot{M}/v_{\mathrm{e}}$) for M- (squares) and S-type AGB stars (triangles). The solid line is a least-squares fit to the data points as described by Eq.~\ref{size_eq}. The dashed and dotted lines show the results from a simple photochemical model (Sect.~\ref{sect_size}). The dashed line is calculated for $h$=0.2 and the dotted line for $h$=0.5 (see text for details). The carbon star R~For (dot) is a also shown in the plots.}
       \label{fig_size}
            \end{figure}

In Fig.~\ref{fig_size} the fractional abundances for this sub-sample are plotted as a function of a density measure ($\dot{M}/v_{\mathrm{e}}$). No apparent trends with density are present. However, the S-type AGB stars appear on average to have larger HCN fractional abundances than M-type AGB stars (this is further discussed in Sect.~\ref{ss:abund}). The carbon star \object{R~For} is again shown for comparison. 

We find that the derived envelope sizes in Table~\ref{table_chi2} scale with density as
\begin{equation}
\label{size_eq}
\log r_{\mathrm{e}} = 19.9 (\pm\,0.6) + 0.55 (\pm\,0.09) \log \left( \frac{\dot{M}}{v_{\mathrm{e}}} \right),
\end{equation}
where $\upsilon_{\mathrm{e}}$ is the expansion velocity of the wind in km\,s$^{-1}$, and $\dot{M}$ the mass-loss rate in M$_{\odot}$\,yr$^{-1}$ (the values in parenthesis are 1\,$\sigma$ errors). As shown in Fig.~\ref{fig_size}, the scaling law gives a good estimate of the envelope sizes, within the uncertainties of the results from the multi-transition radiative transfer. The reduced $\chi^2$ of the fit is 1.5 indicating a good fit. \citet{Delgado03b} derived a slope of 0.48 in their analysis of SiO line emission which is consistent with the value of 0.55 obtained here, within the uncertainties. The HCN sizes derived here are systematically larger than the corresponding SiO envelope sizes, although they are still comparable within the mutual uncertainties of the scaling laws. 

High angular resolution HCN observations, capable of resolving the emitting regions, require the use of interferometers and exist only for a small number of sources. Such observations generally put better constraints on the HCN abundance distribution than multi-transition single-dish observations and in order to test our relation (Eq.~\ref{size_eq}) we have compared it to previous interferometer results. \citet{Lindqvist00} estimated envelope sizes for a sample of five carbon stars with varying mass-loss rates. In three cases it was possible to obtain good constraints. For the carbon stars \object{CW Leo} and \object{LP And}, Eq.~\ref{size_eq} gives estimates that agree very well with the findings of \citet{Lindqvist00}. In the case of RW~LMi, Eq.~\ref{size_eq} gives an envelope size that is about a factor of two too small. 

\begin{figure}
      \centerline{\includegraphics[width=6cm]{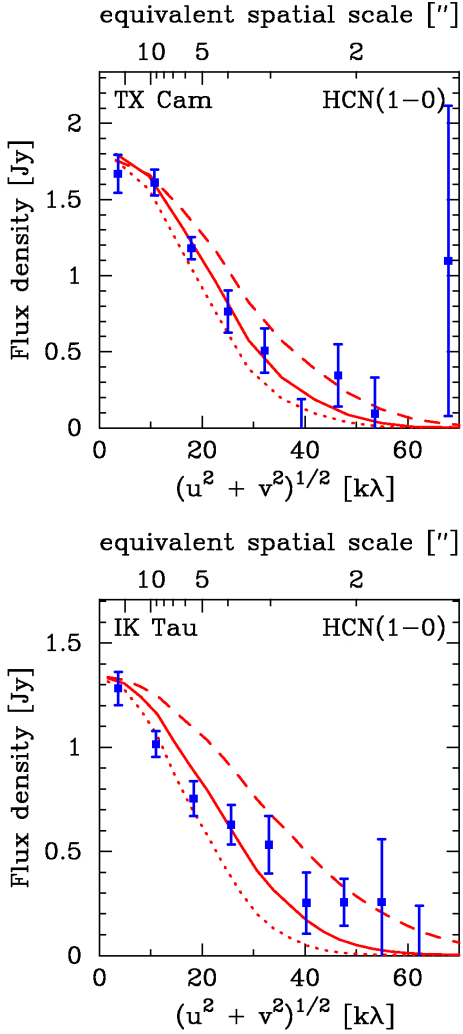}}
   \caption{Visibilities, averaged over 2\,km\,s$^{-1}$ around the systemic velocity, obtained for IK Tau and TX Cam using the Owens Valley Millimeter Array \citep{Marvel05}. The observations are overlaid by model results using various abundance distributions. 
   {\em Top panel --} For TX Cam the following models are shown: $f_0$\,$=$\,1.1\,$\times$\,10$^{-6}$ and $r_{\mathrm{e}}$\,$=$\,1.2\,$\times$\,10$^{16}$\,cm (dotted line), $f_0$\,$=$\,7.0\,$\times$\,10$^{-7}$ and $r_{\mathrm{e}}$\,$=$\,1.6\,$\times$\,10$^{16}$\,cm (solid line), and $f_0$\,$=$\,5.0\,$\times$\,10$^{-7}$ and $r_{\mathrm{e}}$\,$=$\,2.0\,$\times$\,10$^{16}$\,cm (dashed line).
   {\em Bottom panel --} For IK Tau the following models are shown: $f_0$\,$=$\,1.0\,$\times$\,10$^{-6}$ and $r_{\mathrm{e}}$\,$=$\,6.0\,$\times$\,10$^{15}$\,cm (dotted line), $f_0$\,$=$\,5.0\,$\times$\,10$^{-7}$ and $r_{\mathrm{e}}$\,$=$\,9.0\,$\times$\,10$^{15}$\,cm (solid line), and $f_0$\,$=$\,3.0\,$\times$\,10$^{-7}$ and $r_{\mathrm{e}}$\,$=$\,1.2\,$\times$\,10$^{16}$\,cm (dashed line).}
   \label{txcam_iktau}
      \end{figure}

In Fig.~\ref{txcam_iktau} we show the interferometric observations of HCN $J$\,=\,1\,$\rightarrow$\,0 emission towards the two M-type AGB stars TX~Cam and IK~Tau, performed by \citet{Marvel05} at the Owens Valley Radio Observatory (OVRO). Plotted with the data are model results for different HCN abundance distributions within reasonable ranges of the results found in Table~\ref{table_chi2}. The analysis and comparison is performed in the $uv$-plane in order to maximize the sensitivity and resolution of the observations. It is found that envelope sizes of 1.6\,$\times$\,10$^{16}$\,cm for TX~Cam and 9.0\,$\times$\,10$^{15}$\,cm for IK~Tau provide good fits to the observations. These values are consistent with those obtained from the multi-transition analysis of single-dish data reported in Table~\ref{table_chi2}. Furthermore, a comparison between the line strength of the HCN $J$\,=\,1\,$\rightarrow$\,0 emission measured with the OSO 20\,m telescope, and the sum of the emission measured by the OVRO interferometer, shows good agreement, and no indication of extended emission being resolved out by the interferometer. 

Recently, \citet{Muller08} reported interferometric observations of HCN $J$\,=\,8\,$\rightarrow$\,7 emission towards the low-mass-loss-rate M-type AGB star W~Hya. They find, from a detailed radiative transfer analysis, an HCN envelope size of about 3\,$\times$\,10$^{15}$\,cm, in very good agreement with our derived value of 3.0\,$\pm$\,1.1$\times$\,10$^{15}$\,cm (see Table~\ref{table_chi2}). These examples give us confidence in the adopted approach and we feel that the envelope sizes, as estimated by Eq.~\ref{size_eq}, are adequate for use in a statistical study of a large sample of sources such as this.

The HCN envelope sizes can also be estimated using a simple photochemical model. Assuming that the size is regulated by photodissociation \citep{huggglas82}, the abundance distribution is given by:

\begin{equation}
\frac{\mathrm{d} f}{\mathrm{d} r} = \frac{-G_{0} ~ \mathrm{exp}(-\frac{r_{\mathrm{d}}}{r}) }{v_{\mathrm{exp}}}f,
\label{photo}
\end{equation}

where $G_{0}$ is the unshielded photodissociation rate of HCN and $r_{\mathrm{d}}$ is the dust-shielding distance:
\begin{equation}
r_{\mathrm{d}} = 1.4 \frac{3 Q_{\mathrm{abs}}}{4 a_{\mathrm{d}} \rho_{\mathrm{d}}} \frac{\dot{M}_{\mathrm{d}}}{4 \pi v_{\mathrm{d}}}.
\end{equation}
$Q_{\mathrm{abs}}$ is the dust absorption efficiency, $\rho_{\mathrm{d}}$ and $a_{\mathrm{d}}$ are the dust grain density and radius, and $\dot{M}_{\mathrm{d}}$ and $\upsilon_{\mathrm{d}}$ are the dust mass-loss rate and velocity. By using the h-parameter defined in \citet{Schoeier01}, and assuming $\Psi=0.01$, $\rho_{\mathrm{d}}=2$\,g\,cm$^{-3}$, $a_{\mathrm{d}}=0.05$\,$\mu$m, the dust-shielding distance can be re-written as:
\begin{equation}
r_{\mathrm{d}} = 5.27\,10^{22} \, \frac{h \dot{M}}{v_{\mathrm{d}}} \quad{\mathrm{cm}}, \label{sdist2}
\end{equation}
The wavelength dependence of $Q_{\mathrm{abs}}$ in the region of interest, $\sim$1000--3000\,{\AA}, is weak \citep[e.g.,][]{Suh00} and a generic value of 1.0 is adopted. As a comparison to our observational results, the e-folding radius calculated from Eq.~\ref{photo} is shown in Fig.~\ref{fig_size} for $h$\,=\,0.2 (dashed) and $h$\,=\,0.5 (dotted) (typical values for low-, $L\,<\,5000$\,L$_{\odot}$, and high-luminosity, $L\,>\,5000$\,L$_{\odot}$, stars) for different $\dot{M}/v_{\mathrm{e}}$. $G_0$ is taken to be 1.1\,$\times$\,10$^{-9}$\,s$^{-1}$ \citep{vandis88}, and the dust velocity is calculated from the gas velocity, measured from the width of the CO lines, and the drift velocity, calculated from the luminosity and the mass-loss rate \citep{Schoeier01}. The results from the photochemical model is further discussed in Sect.~\ref{chem_mod}.


\subsubsection{HCN maser action}
\label{considerations}
   \begin{figure}
   \centerline{\includegraphics[width=7cm]{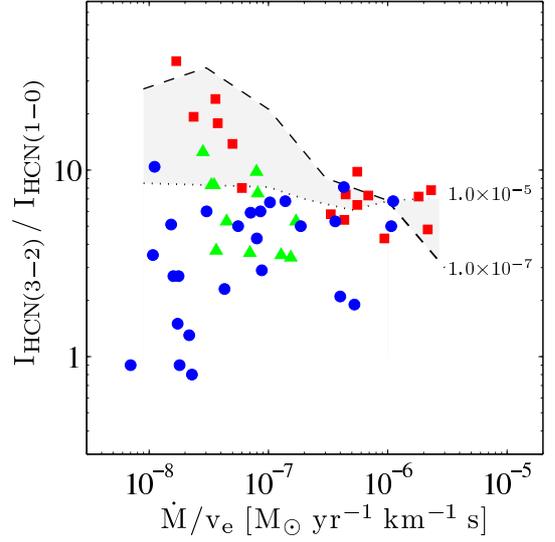}}
   \caption{Observed (markers) and modelled (lines) intensity ratios between the HCN $J$\,$=$\,1\,$\rightarrow$\,0 and $J$\,$=$\,3\,$\rightarrow$\,2 lines as a function of a density measure ($\dot{M}/v_{\mathrm{e}}$) of the CSE. Observations of carbon stars are indicated by blue circles, M-type AGB stars by red squares, and S-type AGB stars by green triangles. Model predictions with HCN abundances of 1\,$\times$\,10$^{-5}$ (solid line) and 1\,$\times$\,10$^{-7}$ (dashed line) are shown for comparison for 
   a circumstellar model as described in the text.}
    \label{ratios_fig}
   \end{figure}
The HCN $J$\,=\,1\,$\rightarrow$\,0  line has been found to exhibit maser features in a number of low-mass carbon stars \citep{Olofsson93b, Olofsson98b}. \citet{Lindqvist00} found that their radiative transfer model could not properly account for the observed intensities in this type of sources. The observed intensities were significantly higher than those predicted by their model. This was further corroborated by \citet{Trung00} in their modelling of the low-mass-loss-rate object Y~CVn. \citet{Lindqvist00} speculated that maser action, emphasized by deviations from a smooth wind, could be the reason for the discrepancy from the model predictions. Recently, \citet{Shinnaga09} have also found evidence for HCN maser emission in the inner wind of the high-mass-loss rate object CW~Leo (IRC\,+10216), suggesting that maser emission is a common phenomenon in carbon stars.

In order to investigate where this phenomenon arises, we have calculated a number of models covering the density range of our sample sources. In these test models the mass-loss rate is varied in the range from 1\,$\times$\,10$^{-7}$\,M$_{\odot}$\,yr$^{-1}$ to 3\,$\times$\,10$^{-5}$\,M$_{\odot}$\,yr$^{-1}$ for a wind that expands at a velocity of 10\,km\,s$^{-1}$. The central star is assumed to have a temperature of 2400\,K and a luminosity of 5000\,L$_{\odot}$ in all models. The CO abundance is assumed to be 6\,$\times$\,10$^{-4}$. To cover the majority of the stars in our sample, two different HCN abundances; 1\,$\times$\,10$^{-5}$ and 1\,$\times$\,10$^{-7}$, were considered. Under certain assumptions, Eq.~\ref{photo} can be solved to give an estimate of the photodissociation radius \citep{Olofsson98b}:
\begin{equation}
\label{olofsson_eq}
r_{\mathrm{ph}} = 10^5\left( \frac{v_{\mathrm e}}{G_0}\right) \mathrm{cm} + 2.9\times 10^{15}\left(\frac{\dot{M}}{10^{-6}}\right)^{0.7} \left(\frac{v_{\mathrm e}}{10}\right)^{-0.4}\mathrm{cm},
\end{equation}
where $\upsilon_{\mathrm{e}}$ is the expansion velocity of the wind in km\,s$^{-1}$, and $\dot{M}$ the mass-loss rate in M$_{\odot}$\,yr$^{-1}$. 
This estimate of the envelope size was then further scaled by a factor of two, as suggested by \citet{Lindqvist00} and \citet{Bieging00} (see also Sect.~\ref{sect_size}). The source was placed at a distance of 500\,pc and the beams for the HCN lines were assumed to correspond to those from a 15\,m telescope, such as the JCMT and the SEST. All observational results shown in Fig.~\ref{ratios_fig} have been scaled to a 15\,m telescope assuming unresolved emission (i.e., that the intensity scales as the inverse of the aperture squared). The uncertainty in the observed intensity ratios is estimated to be $\approx$\,50\%. 

The model line intensity ratios between the $J$\,=\,3\,$\rightarrow$\,2 and $J$\,=\,1\,$\rightarrow$\,0 lines (dashed and dotted lines, one for each assumed HCN abundance) are compared to observed values (markers) in Fig.~\ref{ratios_fig}. There appears to be a distinction between the carbon stars and M-type AGB stars in that carbon stars tend to have lower HCN $J$\,=\,3\,$\rightarrow$\,2 to $J$\,=\,1\,$\rightarrow$\,0 line intensity ratios in the low-density regime. The S-type AGB stars tend to fall in between the two other chemical types. We identify a region where $\dot{M}$/$\upsilon_{\mathrm e}$\,$\lesssim$\,3\,$\times$\,10$^{-8}$\,M$_{\odot}$\,yr$^{-1}$\,km$^{-1}$\,s and the  HCN $J$\,=\,3\,$\rightarrow$\,2 to $J$\,=\,1\,$\rightarrow$\,0 line intensity ratio is $\sim$\,1 for several carbon stars. Within the basic assumptions of the circumstellar model it is not possible to achieve such a low line intensity ratio. We believe these low values are due to maser action (possibly strengthened by a clumpy gas structure) in the $J$\,=\,1\,$\rightarrow$\,0 transition, concluding that under the assumptions of our circumstellar model it will be very difficult to reproduce the observations of the lower-density carbon stars where the conditions for strong maser action are met (high abundance/low density). In the higher density case, the maser action is likely more effectively quenched and therefore the higher-density carbon stars are not subject to the same problems. When comparing model predictions with observations in this case one should keep in mind that a relatively large scatter is expected around the lines representing model results, due to significant deviations in individual sources from these model assumptions. Therefore, Fig.~\ref{ratios_fig} is not giving strict limits for the radiative transfer model, but illustrates the difficulties and emphasizes that specific care has to be taken when modelling and interpreting the results on carbon stars in this region of parameter space. In the following analysis, the modelling of these stars is performed excluding the $J$\,=1\,$\rightarrow$\,0 transition when the (bad) fit to it, dominates the total $\chi^{2}$-estimate.


\subsubsection{HCN abundances}
\label{ss:abund}
\begin{figure}[h]
      \centerline{\includegraphics[width=6cm,angle=-90]{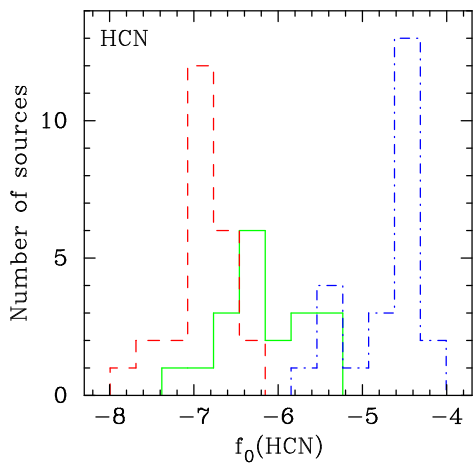}}
   \caption{Histograms showing the derived HCN fractional abundances ($f_0$) for the carbon stars (dash-dotted, blue line; 25 stars),  M-type AGB stars (dashed, red line; 25 stars), and S-type AGB stars (solid, green line; 19 stars).}
     \label{fig_hist}
        \end{figure}

\begin{figure*}[t]
      \centerline{\includegraphics[width=18cm]{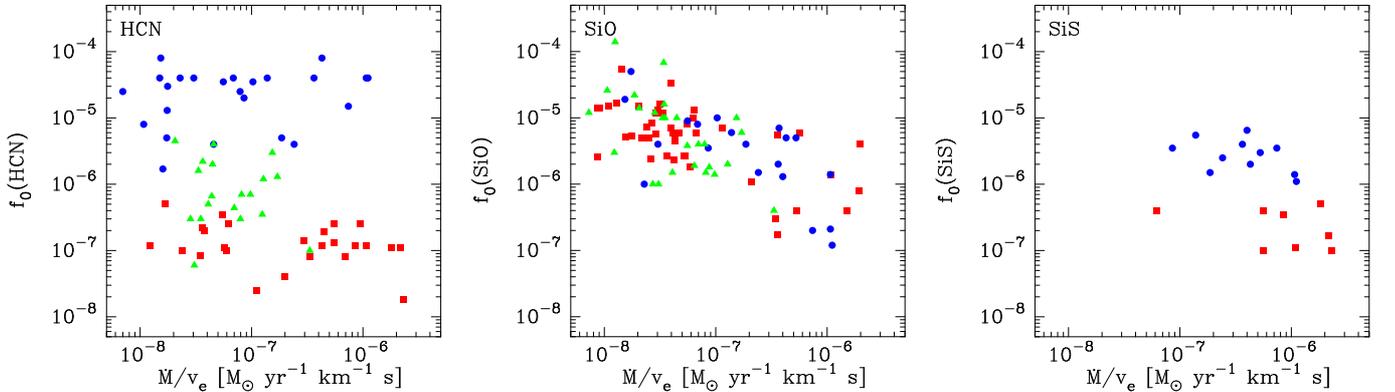}}
   \caption{Results from multi-transition excitation analysis of circumstellar line emission towards a large sample of AGB stars with varying physical and chemical properties. {\bf left --} HCN fractional abundance, $f_0({\mathrm{HCN}})$, as a function of a density measure ($\dot{M}$/$\upsilon_{\mathrm e}$) for carbon stars (filled circles), M-type AGB stars (filled squares) and S-type AGB stars (filled triangles).  {\bf middle --} SiO fractional abundance, $f_0({\mathrm{SiO}})$, as a function of a density measure ($\dot{M}$/$\upsilon_{\mathrm e}$) for carbon stars \citep[filled circles; ][]{Schoeier06a}, M-type AGB stars \citep[filled squares; ][]{Delgado03b} and S-type AGB stars \citep[filled triangles; ][]{Ramstedt09}.  {\bf right --} SiS fractional abundance, $f_0({\mathrm{SiS}})$, as a function of a density measure for carbon stars (filled circles) and M-type AGB stars (filled squares) from \citet{Schoeier07a}.}
     \label{abundance_fig}
        \end{figure*}

Using the scaling relation in Eq.~\ref{size_eq} it is now possible to estimate the HCN envelope sizes for all of the sample sources. Once the size, $r_\mathrm{e}$, is known only one free parameter remains in the excitation analysis, the initial abundance $f_0$. The derived abundances and adopted envelope sizes of HCN are reported in Tables~\ref{sample_c} and \ref{sample_sm}, and Figs~\ref{fig_hist} and \ref{abundance_fig}. The fits to the data are generally good with reduced $\chi^2$ values of $\approx$\,1-3. It is found that the abundance of HCN varies substantially with the photospheric C/O-ratio of the central star. We find that the median HCN abundance is 2.9\,$\times$\,10$^{-5}$ for the carbon stars (25 stars; dash-dotted line in Fig.~\ref{fig_hist} and filled circles in Fig.~\ref{abundance_fig}), more than two orders of magnitudes higher than for the M-type AGB stars (median is 1.2\,$\times$\,10$^{-7}$; 25 stars; dashed line in Fig.~\ref{fig_hist} and filled squares in Fig.~\ref{abundance_fig}). The S-type AGB stars, on the other hand, seem to be distributed in between the two other types (19 stars; solid line in Fig.~\ref{fig_hist} and filled triangles in Fig.~\ref{abundance_fig}) with a median HCN abundance of 7.0\,$\times$\,10$^{-7}$. The spread in abundances (as measured by the ratio of the 75$^{\mathrm{th}}$ and 25$^{\mathrm{th}}$ percentile) is also larger for the S-type AGB stars (5.6) compared to those of the carbon stars (4) and the M-type AGB stars (2), possibly indicating the sensitivity of the HCN abundance in the border area between an oxygen-dominated chemistry (lower HCN abundances) and a carbon-dominated chemistry (higher HCN abundances) within the S-type sample.


\begin{table*}
\caption{Model results.}
\label{sample_c}
$$
\begin{array}{p{0.14\linewidth}cccccccccccccccccccccc}
\hline
\noalign{\smallskip}
& 
\multicolumn{8}{c}{\mathrm{SED\ modelling}} & &
\multicolumn{4}{c}{\mathrm{CO\ modelling}} & &
\multicolumn{4}{c}{\mathrm{HCN\ modelling}} \\
\noalign{\smallskip}
\cline{2-9}\cline{11-14}\cline{16-19}
\noalign{\smallskip}
\multicolumn{1}{c}{{\mathrm{Source}}} &
\multicolumn{1}{c}{D}& 
\multicolumn{1}{c}{L_{\star}}&
\multicolumn{1}{c}{T_{\star}}&
\multicolumn{1}{c}{\tau_{10}}&
\multicolumn{1}{c}{T_{\mathrm{d}}(r_{\mathrm{i}})}&
\multicolumn{1}{c}{r_{\mathrm{i}}}&
\multicolumn{1}{c}{\chi^2_{\mathrm{red}}} &
\multicolumn{1}{c}{N} & &
\multicolumn{1}{c}{\dot{M}}&
\multicolumn{1}{c}{v_{\mathrm{e}}} &
\multicolumn{1}{c}{\chi^2_{\mathrm{red}}} &
\multicolumn{1}{c}{N} & &
\multicolumn{1}{c}{f_0} & 
\multicolumn{1}{c}{r_{\mathrm{e}}} &
\multicolumn{1}{c}{\chi^2_{\mathrm{red}}} &
\multicolumn{1}{c}{N} \\
&
\multicolumn{1}{c}{[\mathrm{pc}]}  &
\multicolumn{1}{c}{[\mathrm{L_{\odot}}]} & 
\multicolumn{1}{c}{[\mathrm{K}]} & &
\multicolumn{1}{c}{[\mathrm{K}]} &
\multicolumn{1}{c}{[\mathrm{cm}]} & & & &
\multicolumn{1}{c}{[\mathrm{M_{\odot}\,yr^{-1}}]} &
\multicolumn{1}{c}{[\mathrm{km\,s^{-1}}]} & & 
&&&
\multicolumn{1}{c}{[\mathrm{cm}]} & &  \\
\noalign{\smallskip}
\hline
\noalign{\smallskip}
\em{Carbon stars} \\
\ \object{LP And} 	&\phantom{0}630 & 9400 & 2000 & 0.60\phantom{0} & 1100 & 1.8\times10^{14}& 0.8 & 11 && 1.5\times10^{-5} & 13.5 & 0.7 & 7  && 4.0\times10^{-5}\phantom{:} & 4.2\times10^{16} & 3.0 & 3  \\
\ \object{V Aql}		&\phantom{0}330 & 6500 & 2800 & 0.02\phantom{0} & 1500 & 6.1\times10^{13}& 1.7 & 10 && 1.4\times10^{-7} & \phantom{0}8.0 & 1.4 & 5  && 1.3\times10^{-5}\phantom{:} & 4.3\times10^{15} & 1.6 & 3  \\
\ \object{RV Aqr}                       & \phantom{0}670 & 6800 & 2200 & 0.27\phantom{0} & 1300 & 7.6\times10^{13} & 0.8 & 9  & & 2.8\times10^{-6} & 15.0 & 0.3 & 3 && 5.0\times10^{-6}\phantom{:} & 1.6\times10^{16} & 2.6 & 3\\
\ \object{S Aur}                          & \phantom{0}300 & 8900 & 3000 & 0.005 & 1500  &  7.3\times10^{13}& 8.1 & 7 && 4.0\times10^{-7} & 24.0 & 1.1 & 3  && 3.0\times10^{-6} & 4.2\times10^{15} & \cdots & 1 \\
\ \object{UU Aur}                       & \phantom{0}260 & 6900 & 2800 & 0.017 & 1500  &  6.3\times10^{13}& 1.3 & 9 && 2.4\times10^{-7} &10.5& 1.0 & 5  && 4.0\times10^{-5} & 5.0\times10^{15} & \cdots & 1 \\
\ \object{ST Cam}                      & \phantom{0}360 & 4400 & 2800 & 0.02\phantom{0} & 1500 & 5.0\times10^{13}& 1.0 & 7 && 1.3\times10^{-7} & \phantom{0}9.5 & 1.1 & 3 && 3.0\times10^{-5}\phantom{:} & 3.8\times10^{15} & 0.1 & 2 \\
\ \object{HV Cas}                      & \phantom{0}970 & 7900 & 2200 & 0.10\phantom{0} & 1500 & 6.4\times10^{13}& 1.0 & 7 && 1.5\times10^{-6} & 19.0 & 1.1 & 3 && 2.5\times10^{-5}\phantom{:} & 9.8\times10^{15} & 0.7 & 3 \\
\ \object{S Cep}                         & \phantom{0}380 & 7300 & 2200 & 0.12\phantom{0} & 1400 & 5.8\times10^{13} & 1.5 & 9 && 1.2\times10^{-6} & 21.5 & 0.9 & 5 && 3.5\times10^{-5}\phantom{:} & 8.2\times10^{15} & 1.4 & 6 \\
\ \object{X Cnc}                         & \phantom{0}280 & 2800 & 2200 & \cdots & \cdots  &  5.0\times10^{13}& \cdots & \cdots && 7.0\times10^{-8} &\phantom{0}6.5& 2.6 & 5  && 8.0\times10^{-6} & 3.3\times10^{15} & \cdots & 1 \\
\ \object{Y CVn}                         & \phantom{0}220 & 4400 & 2200 & 0.01\phantom{0} & 1500 & 8.7\times10^{13}& 0.6 & 8 && 1.5\times10^{-7} & \phantom{0}8.5 & 0.5 & 5 && 3.0\times10^{-5}\phantom{:} & 1.0\times10^{16} & 0.1 & 3 \\
\ \object{V Cyg}                         & \phantom{0}310 & 6300 & 1900 & 0.08\phantom{0} & 1200 & 8.7\times10^{13}& 0.6 & 8 && 9.0\times10^{-7} & 10.5 & 0.5 & 5 && 2.0\times10^{-5}\phantom{:} & 1.0\times10^{16} & 1.6 & 5 \\
\ \object{UX Dra}                       &\phantom{0}570 & 4000 & 2600 & \cdots & \cdots & 1.0\times10^{14}& \cdots & \cdots && 1.6\times10^{-7} & \phantom{0}3.5 & 0.0 & 0  && 5.0\times10^{-6}\phantom{:} & 7.3\times10^{15} & \cdots & 1 \\
\ \object{R For}                           &\phantom{0}610 & 5800 & 2000 & 0.25\phantom{0} & 1400 &  5.6\times10^{13}& 2.8 & 9 && 1.1\times10^{-6} & 15.5 & 1.5 & 7 && 4.0\times10^{-5}\phantom{:} & 9.3\times10^{15} & 0.5 & 4  \\
\ \object{V821 Her}                   & \phantom{0}600 & 7900 & 2200 & 0.45\phantom{0} & 1500 & 8.1\times10^{13}& 2.4 & 10  && 1.8\times10^{-6} & 13.0 & 3.9 & 4  && 4.0\times10^{-5}\phantom{:} & 1.3\times10^{16} & 0.8 & 4 \\
\ \object{U Hya}                          &\phantom{0}160 & 2500 & 2400 & \cdots & \cdots & 1.0\times10^{14}& \cdots & \cdots && 8.0\times10^{-8} & \phantom{0}5.0 & 0.0 & 0  && 1.7\times10^{-6}\phantom{:} & 4.3\times10^{15} & 0.1 & 2 \\
\ \object{CW Leo}                      & \phantom{0}120 & 9600 & 2000 & 0.90\phantom{0} & 1200 & 1.7\times10^{14}& 2.1 & 9 && 1.5\times10^{-5} & 14.0 & 0.5 & 8 && 4.0\times10^{-5}\phantom{:} & 4.1\times10^{16} & 1.0 & 9 \\
\ \object{R Lep}                         & \phantom{0}250 & 4000 & 2200 & 0.06\phantom{0} &  1500 & 4.3\times10^{13}& 0.3 & 9 && 5.0\times10^{-7} & 16.5 & 0.7 & 6 && 1.0\times10^{-5}\phantom{:} & 5.8\times10^{15} & 2.4 & 4  \\
\ \object{RW LMi}                      & \phantom{0}440 & 9700 & 2000 & 0.50\phantom{0} & 1000 & 2.1\times10^{14}& 1.4 & 11 && 6.0\times10^{-6} & 16.5 & 1.1 & 7 && 4.0\times10^{-5}\phantom{:} & 2.3\times10^{16} & 1.5 & 8  \\
\ \object{W Ori}                           &\phantom{0}220 & 3500 & 2600 & 0.02\phantom{0} & 1500 & 4.3\times10^{13}& 1.6 &  8 && 7.0\times10^{-8} & 10.0 & 1.0 & 6 && 2.5\times10^{-5}\phantom{:} & 2.6\times10^{15} & 8.8 & 2  \\
\ \object{V384 Per}                    & \phantom{0}560 & 8100 & 2000 & 0.25\phantom{0} & 1300 & 1.0\times10^{14}& 1.5 & 11  && 3.5\times10^{-6} & 14.5 & 0.7 & 6 && 4.0\times10^{-6}\phantom{:} & 1.8\times10^{16} & 3.1 & 2 \\
\ \object{V466 Per}                    & \phantom{0}350 & 8100 & 2000 & 0.05\phantom{0} & 1500 & 5.6\times10^{13}& 2.0 & 8  && 1.0\times10^{-7} & 9.0 & 1.2 & 3 && 4.0\times10^{-6}\phantom{:} & 3.4\times10^{15} & 3.1 & 2 \\
\ \object{W Pic}                          &   \phantom{0}490  & 4000 & 2500 & 0.015  & 1500  & 4.5\times10^{13} &  3.2  & 8  && 2.3\times10^{-7} & 15.0& 3.4 & 3  && 5.0\times10^{-5} & 4.0\times10^{15} & 3.5 & 2 \\
\ \object{X TrA}                          &\phantom{0}230 & 4000 & 2200 & \cdots & \cdots & 1.0\times10^{14}& \cdots & \cdots && 1.3\times10^{-7} & \phantom{0}7.5 & 0.0 & 0  && 5.0\times10^{-6}\phantom{:} & 4.3\times10^{15} & 0.1 & 2 \\
\ \object{R Vol}                           &\phantom{0}730 & 4000 & 2200 & 0.30\phantom{0} & 1500 & 6.6\times10^{13}& 1.1 &  9 && 1.7\times10^{-6} & 16.5 & 0.6 & 3 && 3.5\times10^{-5}\phantom{:} & 7.0\times10^{15} & 0.1 & 2  \\
\ \object{AFGL 3068}                & \phantom{0}980 & 7800 & 2000 & 2.70\phantom{0} & 1100 & 2.5\times10^{14}& 1.8 & 8  && 1.0\times10^{-5} & 13.5 & 0.6 & 4  && 1.5\times10^{-5}\phantom{:} & 3.4\times10^{16} & 3.1 & 4 \\
\ \object{IRAS\,15194--5115}  & \phantom{0}500 & 8800 & 2400 & 0.55\phantom{0} & 1200 & 1.5\times10^{14}& 0.4 &  9 && 9.0\times10^{-6} & 21.0 & 0.9 & 4   && 8.0\times10^{-5}\phantom{:} & 2.5\times10^{16} & 5.0 & 4 \\
\noalign{\smallskip}
\hline
\end{array}
$$
\noindent
\end{table*}

\begin{table*}
\caption{Model results.}
\label{sample_sm}
$$
\begin{array}{p{0.14\linewidth}cccccccccccccccccccccc}
\hline
\noalign{\smallskip}
& 
\multicolumn{8}{c}{\mathrm{SED\ modelling}} & &
\multicolumn{4}{c}{\mathrm{CO\ modelling}} & &
\multicolumn{4}{c}{\mathrm{HCN\ modelling}} \\
\noalign{\smallskip}
\cline{2-9}\cline{11-14}\cline{16-19}
\noalign{\smallskip}
\multicolumn{1}{c}{{\mathrm{Source}}} &
\multicolumn{1}{c}{D}& 
\multicolumn{1}{c}{L_{\star}}&
\multicolumn{1}{c}{T_{\star}}&
\multicolumn{1}{c}{\tau_{10}}&
\multicolumn{1}{c}{T_{\mathrm{d}}(r_{\mathrm{i}})}&
\multicolumn{1}{c}{r_{\mathrm{i}}}&
\multicolumn{1}{c}{\chi^2_{\mathrm{red}}} &
\multicolumn{1}{c}{N} & &
\multicolumn{1}{c}{\dot{M}}&
\multicolumn{1}{c}{v_{\mathrm{e}}} &
\multicolumn{1}{c}{\chi^2_{\mathrm{red}}} &
\multicolumn{1}{c}{N} & &
\multicolumn{1}{c}{f_0} & 
\multicolumn{1}{c}{r_{\mathrm{e}}} &
\multicolumn{1}{c}{\chi^2_{\mathrm{red}}} &
\multicolumn{1}{c}{N} \\
&
\multicolumn{1}{c}{[\mathrm{pc}]}  &
\multicolumn{1}{c}{[\mathrm{L_{\odot}}]} & 
\multicolumn{1}{c}{[\mathrm{K}]} & &
\multicolumn{1}{c}{[\mathrm{K}]} &
\multicolumn{1}{c}{[\mathrm{cm}]} & & & &
\multicolumn{1}{c}{[\mathrm{M_{\odot}\,yr^{-1}}]} &
\multicolumn{1}{c}{[\mathrm{km\,s^{-1}}]} & & 
&&&
\multicolumn{1}{c}{[\mathrm{cm}]} & &  \\
\noalign{\smallskip}
\hline
\noalign{\smallskip}

\em{S-type stars}\\
\ \object{R And}              &\phantom{0}300 & 6000 & 1800 & 0.02\phantom{0} & \phantom{0}500 & 4.9\times10^{14} & 1.9 & 7 && 6.6\times10^{-7} & \phantom{0}8.3 & 1.3  &  5  && \phantom{<:}3.0\times10^{-7} & 9.9\times10^{15} & 3.2 & 6  \\
\ \object{W And}              &\phantom{0}280 & 5800 & 2400 & \cdots & \phantom{0}\cdots & 1.5\times10^{14} & \cdots & \cdots && 1.7\times10^{-7} & \phantom{0}6.0 & 1.1  &  5  && \phantom{<:}3.0\times10^{-7} & 5.6\times10^{15} & 1.3 & 3  \\
\ \object{W Aql}              &\phantom{0}230 & 6800 & 1800 & 0.10\phantom{0} & \phantom{0}600 & 3.7\times10^{14} & 2.2 & 7 && 22\times10^{-7} & 17.2 & 0.6  & 5  && \phantom{<:}1.2\times10^{-6} & 1.3\times10^{16} & 2.4 & 7  \\
\ \object{S Cas}              &\phantom{0}440 & 8000 & 1800 & 0.10\phantom{0} & \phantom{0}700 & 3.7\times10^{14} & 2.9 & 7 && 35\times10^{-7} & 20.5 & 3.6  & 4  && \phantom{<:}1.3\times10^{-6} & 1.5\times10^{16} & 0.7 & 4  \\
\ \object{WY Cas}              &\phantom{0}600 & 6700 & 2200 & \cdots & \phantom{0}\cdots & 2.0\times10^{14} & \cdots & \cdots && 11\times10^{-7} & 13.5 & 6.6 &  3  && \phantom{<:}7.0\times10^{-7} & 1.0\times10^{16} & 0.7 & 2  \\
\ \object{TT Cen}              &\phantom{0}880 & 6500 & 2400 & \cdots & \phantom{0}\cdots & 1.7\times10^{14} & \cdots & \cdots && 25\times10^{-7} & \phantom{0}20.0 & 7.4  &  2  && \phantom{<:}3.5\times10^{-7} & 1.3\times10^{16} & \cdots & 1  \\
\ \object{R Cyg}              &\phantom{0}440 & 6100 & 2200 & \cdots & \phantom{0}\cdots & 1.9\times10^{14} & \cdots & \cdots && 6.3\times10^{-7} & \phantom{0}9.0 & 5.0  &  5  && \phantom{<:}4.4\times10^{-7} & 9.2\times10^{15} & 5.5 & 4  \\
\ \object{$\chi$ Cyg}              &\phantom{0}110 & 5900 & 2400 & 0.03\phantom{0} & 1400 & 0.6\times10^{14} & 3.6 & 7 && 3.8\times10^{-7} & \phantom{0}8.5 & 2.0  &  6  && \phantom{<:}2.0\times10^{-6} & 7.2\times10^{15} & 1.6 & 4  \\
\ \object{R Gem}              &\phantom{0}700 & 5500 & 2400 & \cdots & \phantom{0}\cdots & 1.5\times10^{14} & \cdots & \cdots && 4.4\times10^{-7} & \phantom{0}4.5 & 3.7  & 4  && \phantom{<:}7.0\times10^{-7} & 1.1\times10^{16} & \cdots & 1  \\
\ \object{R Lyn}              &\phantom{0}850 & 5600 & 2400 & \cdots & \phantom{0}\cdots & 1.5\times10^{14} & \cdots & \cdots && 3.3\times10^{-7} & \phantom{0}7.5 & 0.5  & 5  && \phantom{<:}6.6\times10^{-7} & 7.1\times10^{15} & \cdots & 1  \\
\ \object{Y Lyn}              &\phantom{0}260 & 4000 & 2800 & \cdots & \phantom{0}\cdots & 1.0\times10^{14} & \cdots & \cdots && 2.3\times10^{-7} & \phantom{0}7.5 & 3.9  &  3  && \phantom{<:}6.0\times10^{-8} & 5.9\times10^{15} & \cdots & 1  \\
\ \object{S Lyr}              & 1210 & 6300 & 1800 & 0.09\phantom{0} & \phantom{0}500 & 5.1\times10^{14} & 4.9 & 6 && 20\times10^{-7} & 13.0 & 7.6  &  2  && \phantom{<:}3.0\times10^{-6} & 1.4\times10^{16} & 1.8 & 3  \\
\ \object{RZ Peg}              &\phantom{0}970 & 6300 & 2400 & \cdots & \phantom{0}\cdots & 1.3\times10^{14} & \cdots & \cdots && 4.6\times10^{-7} &12.6 & \cdots  &  1  && \phantom{<:}2.2\times10^{-6} & 6.4\times10^{15} & 7.6 & 2  \\
\ \object{RT Sco}              &\phantom{0}270 & 6400 & 2400 & \cdots & \phantom{0}\cdots & 1.7\times10^{14} & \cdots & \cdots && 4.5\times10^{-7} & 11.0 & 3.2 &  2  && \phantom{<:}5.0\times10^{-7} & 6.9\times10^{15} & 2.9 & 2  \\
\ \object{RZ Sgr}              &\phantom{0}730 & 4000 & 2400 & \cdots & \phantom{0}\cdots & 1.3\times10^{14} & \cdots & \cdots && 30\times10^{-7} & \phantom{0}9.0 & 1.4  &  3  && \phantom{<:}1.0\times10^{-7} & 2.2\times10^{16} & \cdots & 1  \\
\ \object{ST Sgr}              &\phantom{0}540 & 5800 & 2400 & \cdots & \phantom{0}\cdots & 1.6\times10^{14} & \cdots & \cdots && 2.0\times10^{-7} & \phantom{0}6.0 & 0.1  &  2  && \phantom{<:}1.6\times10^{-6} & 6.1\times10^{15} & 1.1 & 3  \\
\ \object{EP Vul}              &\phantom{0}510 & 4000 & 2400 & \cdots & \phantom{0}\cdots & 1.3\times10^{14} & \cdots & \cdots && 2.0\times10^{-7} & \phantom{0}5.7 & 8.9  & 3  && \phantom{<:}3.0\times10^{-7} & 6.3\times10^{15} & \cdots & 1  \\
\ \object{IRC-10401}              &\phantom{0}430 & 4000 & 1800 & 0.07\phantom{0} & \phantom{0}500 & 4.1\times10^{14} & 2.9 & 6 && 3.5\times10^{-7} & 17.0 & 13.4  &  2  && \phantom{<:}4.5\times10^{-6} & 4.7\times10^{15} & \cdots & 1  \\

\em{M-type stars} \\ 
\ \object{RR Aql}                 &\phantom{0}530 & \phantom{0}7900 & 2000 & 0.70\phantom{0} & 1500 & 5.9\times10^{13} & 0.8 & 7 && 1.4\times10^{-6} & \phantom{0}7.0 & 0.3  &  3  && \phantom{<:}4.0\times10^{-8} & 1.6\times10^{16} & \cdots & 1  \\
\ \object{RX Boo}                &\phantom{0}120 & \phantom{0}4000 & 1800 & 0.20\phantom{0} & \phantom{0}900 & 1.5\times10^{14} & 0.3 & 8 && 3.2\times10^{-7} & \phantom{0}8.5 & 1.0  &  2  && \phantom{<:}2.0\times10^{-7} & 6.6\times10^{15} & 4.4 & 6  \\
\ \object{BX Cam}               &\phantom{0}500 & \phantom{0}7500 & 2800 & 1.30\phantom{0} & 1500 & 7.1\times10^{13}& 0.2 & 7 && 8.0\times10^{-6} & 18.0 & 2.3 & 3  && \phantom{<:}1.9\times10^{-7} & 2.5\times10^{16} & 1.2 & 3  \\
\ \object{TX Cam}               &\phantom{0}380 & \phantom{0}8600 & 2600 & 1.00\phantom{0} & 1300 & 1.0\times10^{14}& 1.1 & 8 && 1.0\times10^{-5} & 18.0 & 1.0 & 5  && \phantom{<:}2.5\times10^{-7} & 2.9\times10^{16} & 0.7 & 7  \\
\ \object{R Cas}                  &\phantom{0}110 & \phantom{0}3500 & 2800 & 0.06\phantom{0} & \phantom{0}900 & 1.4\times10^{14}& 0.3 & 8 && 5.0\times10^{-7} & 10.0 & 2.0 & 4  && \phantom{<:}2.5\times10^{-7} & 7.7\times10^{15} & 3.2 & 6  \\
\ \object{R Crt}                    &\phantom{0}170 & \phantom{0}4000 & 2800 & 0.03\phantom{0} & \phantom{0}600 & 3.5\times10^{14}& 0.4 & 9 && 5.5\times10^{-7} & 10.0 & 2.2 & 4  && \phantom{<:}3.5\times10^{-7} & 8.1\times10^{15} & 0.1 & 2  \\
\ \object{R Dor}                    &\phantom{00}45 & \phantom{0}4000 & 2100 & 0.03\phantom{0} & 1200 & 9.8\times10^{13}& 0.3 & 9 && 1.3\times10^{-7} & \phantom{0}5.5 & 2.2 & 5  && \phantom{<:}1.0\times10^{-7} & 5.1\times10^{15} & 1.1 & 3  \\
\ \object{R Hor}                  &\phantom{0}310   & \phantom{0}8500 & 2200 & 0.30\phantom{0} & 1400 & 7.2\times10^{13}& 0.9 & 7 && 5.5\times10^{-7} & \phantom{0}5.0 & \cdots & 1  && \phantom{<:}1.0\times10^{-7} & 1.2\times10^{16} & 0.1 & 1  \\
\ \object{R Hya}                  &\phantom{0}150   & \phantom{0}7700 & 2600 & 0.15\phantom{0} & 1500 & 6.1\times10^{13}& 0.4 & 9 && 2.1\times10^{-7} & \phantom{0}6.0 & 5.4 & 3  && \phantom{<:}8.5\times10^{-8} & 6.3\times10^{15} & \cdots & 1  \\
\ \object{W Hya}                  &\phantom{00}78   & \phantom{0}5400 & 2600 & 0.06\phantom{0} & \phantom{0}900 & 8.7\times10^{13}& 0.3 & 8 && 1.1\times10^{-7} & \phantom{0}6.5 & 2.1 & 4  && \phantom{<:}5.0\times10^{-7} & 4.2\times10^{15} & 3.1 & 4  \\
\ \object{R Leo}                  &\phantom{0}130 & \phantom{0}4600 & 1800 & 0.10\phantom{0} & 1100 & 8.2\times10^{13}& 2.7 & 8 && 1.8\times10^{-7} & \phantom{0}5.0 & 0.5 & 3  && \phantom{<:}2.2\times10^{-7} & 6.8\times10^{15} & 0.6 & 4  \\
\ \object{AP Lyn}                  &\phantom{0}420 & \phantom{0}4000 & 2200 & 0.50\phantom{0} & 1000 & 1.1\times10^{14}& 1.5 & 7 && 6.5\times10^{-6} & 15.0 & 0.3 & 3  && \phantom{<:}1.2\times10^{-7} & 2.5\times10^{16} & 0.3 & 2  \\
\ \object{GX Mon}               &\phantom{0}550 & \phantom{0}8200 & 2600 & 2.00\phantom{0} & \phantom{0}900 & 1.1\times10^{14}& 5.0 & 9 && 2.0\times10^{-5} & 18.5 & 2.0 & 4  && \phantom{<:}1.2\times10^{-7} & 4.2\times10^{16} & 0.9 & 2  \\
\ \object{WX Psc}                &\phantom{0}700 & 10300 & 1800 & 3.00\phantom{0} & 1000 & 1.1\times10^{14}& 1.3 & 9 && 4.0\times10^{-5} & 18.5 & 2.0 & 4  && \phantom{<:}1.1\times10^{-7} & 6.1\times10^{16} & 0.8 & 3  \\
\ \object{L$^2$ Pup}          &\phantom{00}86 & \phantom{0}4000 & 2600 & \cdots & \cdots & 1.0\times10^{14}& 1.0 & 7 && 2.7\times10^{-8} & \phantom{0}2.2 & 0.8 & 4  && \phantom{<:}1.2\times10^{-7} & 3.5\times10^{15} & \cdots & 1  \\
\ \object{IK Tau}                  &\phantom{0}260 & \phantom{0}7700 & 2400 & 2.00\phantom{0} & 1500 & 7.8\times10^{13}& 1.0 & 9 && 1.0\times10^{-5} & 18.0 & 0.7 & 4  && \phantom{<:}1.3\times10^{-7} & 2.9\times10^{16} & 2.4 & 4  \\
\ \object{RT Vir}                  &\phantom{0}170 & \phantom{0}4000 & 2800 & 0.03\phantom{0} & \phantom{0}700 & 2.6\times10^{14} & 0.3 & 9 && 4.5\times10^{-7} & \phantom{0}7.5 & 0.9 & 4  && \phantom{<:}1.0\times10^{-7} & 8.5\times10^{15} & 0.3 & 2  \\
\ \object{SW Vir}                  &\phantom{0}170 & \phantom{0}4000 & 2400 & 0.05\phantom{0} & \phantom{0}800 & 1.8\times10^{14} & 0.8 & 9 && 4.0\times10^{-7} & \phantom{0}7.0 & 0.7 & 6  && \phantom{<:}1.1\times10^{-7} & 8.2\times10^{15} & 2.5 & 3  \\
\ \object{CIT\,4}                   &\phantom{0}400 & \phantom{0}4000 & 2200 & 1.30\phantom{0} & 1500 & 4.8\times10^{13} & 0.9 & 7 && 6.0\times10^{-6} & 18.0 & \cdots & 1  && \phantom{<:}8.0\times10^{-8} & 2.2\times10^{16} & 1.0 & 2  \\
\ \object{IRC--10529}        &\phantom{0}620 & 10600 & 2000 & 3.50\phantom{0} & 1100 & 1.8\times10^{14}& 2.2 & 8 && 3.0\times10^{-5} & 13.0 & 4.0 & 5  && \phantom{<:}1.8\times10^{-8} & 6.3\times10^{16} & 3.0 & 5  \\
\ \object{IRC--20197}        &\phantom{0}850 & 14300 & 2200 & 2.00\phantom{0} & 1300 & 1.3\times10^{14}& 0.6 & 7 && 3.7\times10^{-6} & 12.5 & 6.3 & 3  && \phantom{<:}1.4\times10^{-7} & 2.0\times10^{16} & \cdots & 1  \\
\ \object{IRC--30398}        &\phantom{0}600 & \phantom{0}8700 & 2200 & 1.20\phantom{0} & 1200 & 1.2\times10^{14}& 0.4 & 7 && 1.1\times10^{-5} & 16.0 & 0.4 & 2  && \phantom{<:}8.0\times10^{-8} & 3.2\times10^{16} & 1.0 & 4  \\
\ \object{IRC+10365}        &\phantom{0}750 & \phantom{0}7500 & 2000 & 0.75\phantom{0} & \phantom{0}800 & 2.4\times10^{14}& 2.3 & 8 && 1.5\times10^{-5} & 16.0 & 0.1 & 3  && \phantom{<:}2.5\times10^{-7} & 3.8\times10^{16} & 1.8 & 3  \\
\ \object{IRC+40004}         &\phantom{0}680 & 11800 & 2000 & 0.90\phantom{0} & 1100 & 1.5\times10^{14}& 0.9 & 7 && 1.5\times10^{-5} & 17.5 & 0.9 & 2  && \phantom{<:}1.2\times10^{-7} & 3.7\times10^{16} & 3.9 & 2  \\
\ \object{IRC+50137}         & 1200 & \phantom{0}9800 & 2000 & 3.50\phantom{0} & 1100 & 1.7\times10^{14}& 1.2 & 7 && 3.0\times10^{-5} & 16.5 & 1.0 & 2  && \phantom{<:}1.1\times10^{-7} & 5.5\times10^{16} & 0.3 & 2  \\
\noalign{\smallskip}
\hline
\end{array}
$$
\noindent
\end{table*}


There is no apparent trend that the HCN abundance decreases as the density of the wind ($\dot{M}$/$\upsilon_{\mathrm e}$) increases, as would be expected if condensation of HCN molecules onto dust grains were an important process in the dense inner parts of the wind. 
The Pearson correlation coefficients are $r$\,=\,$-$\,0.03 for the carbon stars, $r$\,=\,0.49 for the M-type AGB stars and $r$\,=\,$-$\,0.36 for the S-type AGB stars.
This differs from the result of our previous analysis of SiO \citep{Delgado03b, Schoeier06a, Ramstedt09} where clear correlations with the density of the wind are present. \citet{Schoeier07a} also found weak correlations of SiS fractional abundances with $\dot{M}$/$\upsilon_{\mathrm e}$ for a sample of carbon stars and M-type AGB stars.

The uncertainty in the derived abundances is about a factor of 3 within the adopted model (see Table~\ref{table_chi2}). Additional uncertainties are introduced by the simplifications of the circumstellar model. A systematic error due to the adopted CO/H$_{2}$ abundance might also be present, but we estimate this to be no larger than a factor of $\sim$1.5 \citep{Ramstedt08}. The estimated uncertainties cannot account for the differences that we find between the chemical types. Also, we believe that the spread among the S-type stars is real and indicative of the dependence of the HCN abundance on the C/O-ratio of the star.

The HCN abundances that we obtain agree quite well, within a factor of four, with those presented by \citet{Woods03} using a simple excitation model, for the five high-mass-loss-rate carbon stars in common. The abundances derived in the present analysis are usually higher than the ones reported by \citet{Woods03}.  This discrepancy may be a result of their adopted excitation temperature being too high and of optical depth effects. \citet{Bujarrabal94} found HCN abundances that on average are $\sim$\,20 times higher for the carbon stars than for the M-type AGB stars, supported by the findings in this paper. The average abundances that \citet{Bujarrabal94} derived, through a simple analysis, are 7.4\,$\times$\,10$^{-6}$ for carbon stars, 3.7\,$\times$\,10$^{-7}$ for M-type AGB stars and 4.0\,$\times$\,10$^{-6}$ (only three sources) for S-type AGB stars. \citet{ziuretal09} derived HCN abundances for five oxygen-rich stars using new NRAO data, together with literature data and a model utilizing the Sobolev approximation. For W~Hya the results agree very well, but for TX~Cam and IK~Tau, the differences are too big to be easily explained. Since both sources are high-mass-loss-rate sources it is possible that the discrepancy is due to optical depth effects. For \citet{Lindqvist00} modelled, spatially resolved, interferometric observations of HCN $J$\,=\,1\,$\rightarrow$\,0 line emission from a sample of five carbon stars and derived abundances that all agree within a factor of two with our results for the corresponding sources.

\subsection{Comparison with chemical models; HCN abundances}
\label{chem_mod}
The abundance of HCN derived using a chemical model assuming thermal equilibrium (TE) is highly dependent on the assumed C/O-ratio of the gas. For a typical M-type star (IK Tau with C/O\,=\,0.75, \citet{duaretal99}) the expected TE abundance is more than six orders of magnitude smaller than that derived for a typical carbon star [IRC+10216 with C/O\,=\,1.5, \citet{willcher98}]. For S-type stars (0.98\,$<$\,C/O\,$<$\,1.01) the HCN abundance is expected to have a spread of about three orders of magnitude assuming TE \citep{Cherchneff06}.

The inclusion of shocks in the chemical models drastically alters the expected abundance of HCN, particularly for the oxygen-rich case \citep{Cherchneff06}. For a C/O-ratio of 0.75, the abundance is changed from 2$\times$10$^{-11}$ to 1$\times$10$^{-5}$ at 5\,R$_{\star}$ when shocks are included. For C/O=1.5, the HCN abundance is expected to be lower (3$\times$10$^{-6}$) in the shock model than both the TE value (5$\times$10$^{-5}$) and that derived for the M-type stars. For the S-type stars the spread in the abundances when varying the C/O-ratio around 1 is less than a factor of two, and for C/O=1 an abundance of about 6.5$\times$10$^{-6}$ is expected at 5\,R$_{\star}$.

The derived circumstellar abundances in this work are, on average, slightly less than four orders of magnitude larger for the M-type stars than the expectations from TE models, while the results for the carbon stars agree quite well with TE expectations. The spread in derived abundances around the average is small for both types of stars. Compared to the results from shock chemistry models, our derived abundances are on average about an order of magnitude larger for the carbon stars. A discrepancy of the same order has previously been noted for IRC+10216 \citep{willcher98}. For the M-type stars our results are, on average, approximately one order of magnitude less than the theoretical value from the models including shocks \citep{Cherchneff06}. For the S-type stars, the derived abundances are, on average, about an order of magnitude larger than the TE value. Notably, for these stars, the spread in circumstellar abundances is relatively large. As discussed below (Sect.~\ref{obs_ab}), abundances derived from chemical models including shocks are sensitive to the assumed shock parameters, i.e., the shock velocity \citep[e.g.][]{duaretal99}. Since the observed sources are expected to cover a large range in shock parameters, while the models are often run for one set of parameters, this can sometimes limit the comparison. However, the clear dependence of the circumstellar HCN abundances on the C/O-ratio is in contrast with the expectations from chemical models including shocks, where a homogeneous chemistry is derived with respect to the C/O-ratio. Also, the observational HCN abundances show no, or a very weak, dependence on the wind density, again indicating that the chemistry is not influenced by shocks. Our results are also in contrast with the conclusions of \citet{decietal08} where chemical homogeneity (regardless of C/O-ratio and evolutionary stage) is deduced from observational data (two carbon stars, one S-type star, and one M-type star).

The reason for the discrepancy between the chemical models and the observational results is difficult to identify with any certainty. Given the high abundances found for the M-type stars, non-equilibrium chemical processes (such as shocks) certainly seem to play a role in the formation of HCN. A common formation route for HCN is through the reaction between CN and H$_{2}$ forming HCN and H \citep{duaretal99}. The rate for this reaction is highly temperature dependent and therefore the abundance of HCN found in the chemical models will be very sensitive to the assumed shock properties. However, this would not explain the differences between our results for the different chemistries and the results from chemical models when the same shock model is used \citep{Cherchneff06}. Given that CN is important for the formation of both HCN and CS, a similar investigation as the one presented here, of the abundances of CN and CS would be very helpful for the further interpretation of our results. A complicating factor could be, however, that in the CSEs, CN is also produced through photodissociation of HCN. Previous observations indicate that the abundance of CS is also highly dependent on the C/O-ratio \citep{Olofsson98b}.

\subsection{Observational circumstellar abundances}
\label{obs_ab}

In Fig.~\ref{abundance_fig}, we present the HCN results together with the results of similar abundance analysis on the same samples for SiO \citep{Schoeier06a,Ramstedt09}, and SiS \citep{Schoeier07a}. The results are shown as functions of the circumstellar density. The SiS analysis has only been performed on a sub-sample of carbon and M-type stars.

It is clear that the formation of both HCN and SiS are dependent on the C/O-ratio of the star, although less than what would be expected from TE models. The HCN results are also consistent with the the S-stars constituting a distinct sample chemically. From the theoretical models, SiS is expected to have a dependence on the C/O-ratio, also when chemical non-equilibrium processes are taken into account. SiO, on the other hand, shows no apparent dependence on the C/O-ratio, neither observationally, nor theoretically. According to current chemical models, HCN and SiO are expected to show the same behaviour, and the reason why HCN is significantly more sensitive to the C/O-ratio is not clear. 

In a shock-induced chemistry, the resulting abundances of all molecules are expected to be sensitive to the assumed shock velocity \citep{Cherchneff11}. Although the exact dependence is not entirely straight-forward, a correlation with the pulsation period of the star might therefore be expected. However, no such dependences are found for HCN and SiS, and the one found for SiO probably reflects the correlation between mass-loss rate and pulsation period. This attempt to correlate the estimated abundances with the shock properties is rather crude, because any radial dependence will be smeared out. There are also observational indications that the abundance distribution of all three molecules is more complicated than assumed in our radiative transfer model for many sources \citep[see e.g., line-profile shapes in Fig.~\ref{txcam_spectra}, and][]{Schoeier06c,Schoeier07a,schoetal11}. 

The theoretically predicted abundances do not take condensation onto grains into account, while the observational results apply to the post-condensation region. Hence a comparison between them may yield some insight into dust grain formation and evolution. For SiO there is a clear dependence on the circumstellar density and this is interpreted as due to SiO adsorption onto dust in the high-density wind. On the contrary, the HCN and the SiS abundances show no, or only very weak, dependence on the circumstellar density, in the HCN case over more than two orders of magnitude in density. From this one may conclude that these species are not affected by grain formation and/or evolution. However, we caution that a comparison between the theoretical and observational values is hampered by the significant uncertainties in both models, and by the fact that the chemical theoretical models are usually calculated for only one set of stellar parameter, while the observed samples cover large ranges of stellar mass, luminosity, periods, shock properties, and temperatures.   


\section{Conclusions}
New multi-transition (sub-)millimetre HCN line observations of a statistically significant sample of AGB stars with varying photospheric C/O-ratios and circumstellar properties
are presented. From a detailed excitation analysis, based on a reliable physical model 
of the sources, we reach the following conclusions: 

\begin{itemize}

\item The median derived abundances of HCN are 2.5\,$\times$\,10$^{-5}$, 7.0\,$\times$\,10$^{-7}$, and 1.2\,$\times$\,10$^{-7}$ for the carbon stars (25 stars), S-type AGB stars (19 stars) and M-type AGB stars (25 stars), respectively.

\item We derive a scaling law for the size of the HCN molecular envelope as a function of the wind density ($\dot{M}/v_{\rm{e}}$). The envelope sizes estimated using this law are found to agree with previous interferometric observations and with a simple photochemical model. The HCN envelopes are systematically larger than the corresponding SiO envelopes \citep{Delgado03b}, however still within the mutual uncertainties of the scaling laws.

\item There is a clear dependence of the derived circumstellar HCN abundances on the C/O-ratio of the star. Carbon stars have, on average, about two orders of magnitude higher abundances than M-type AGB stars. The S-type AGB stars, which appear to have a larger spread in their derived HCN abundances, typically fall in between the two other types, however, slightly biased towards the M-type AGB stars. 

\item The sensitivity of the circumstellar HCN abundance to the photospheric C/O-ratio is in stark contrast to predictions from non-LTE chemical models \citep{Cherchneff06}, where very little difference is expected in the HCN fractional abundances between the various types of AGB stars.

\end{itemize}


\begin{acknowledgements}
Fredrik Sch\"oier passed away on the 14th of January 2011. The co-authors have finished this article for him. Fredrik will always be in our hearts and minds. 

FLS, HO and ML are grateful to The Swedish Research council for financial support. SR acknowledges support by the Deutsche Forschungsgemeinschaft (DFG) through the Emmy Noether Research grant VL 61/3-1. This effort/activity is supported by the European Community Framework Programme 7, Advanced Radio Astronomy in Europe, RadioNet grant agreement no.: 227290
\end{acknowledgements}


\bibliographystyle{aa}
\bibliography{AGB,Radtransf,Moldata,Defs}

\clearpage

\clearpage
\pagebreak


\appendix
\section{New observations of HCN line emission}
\begin{figure*}
\centerline{\includegraphics[width=14cm]{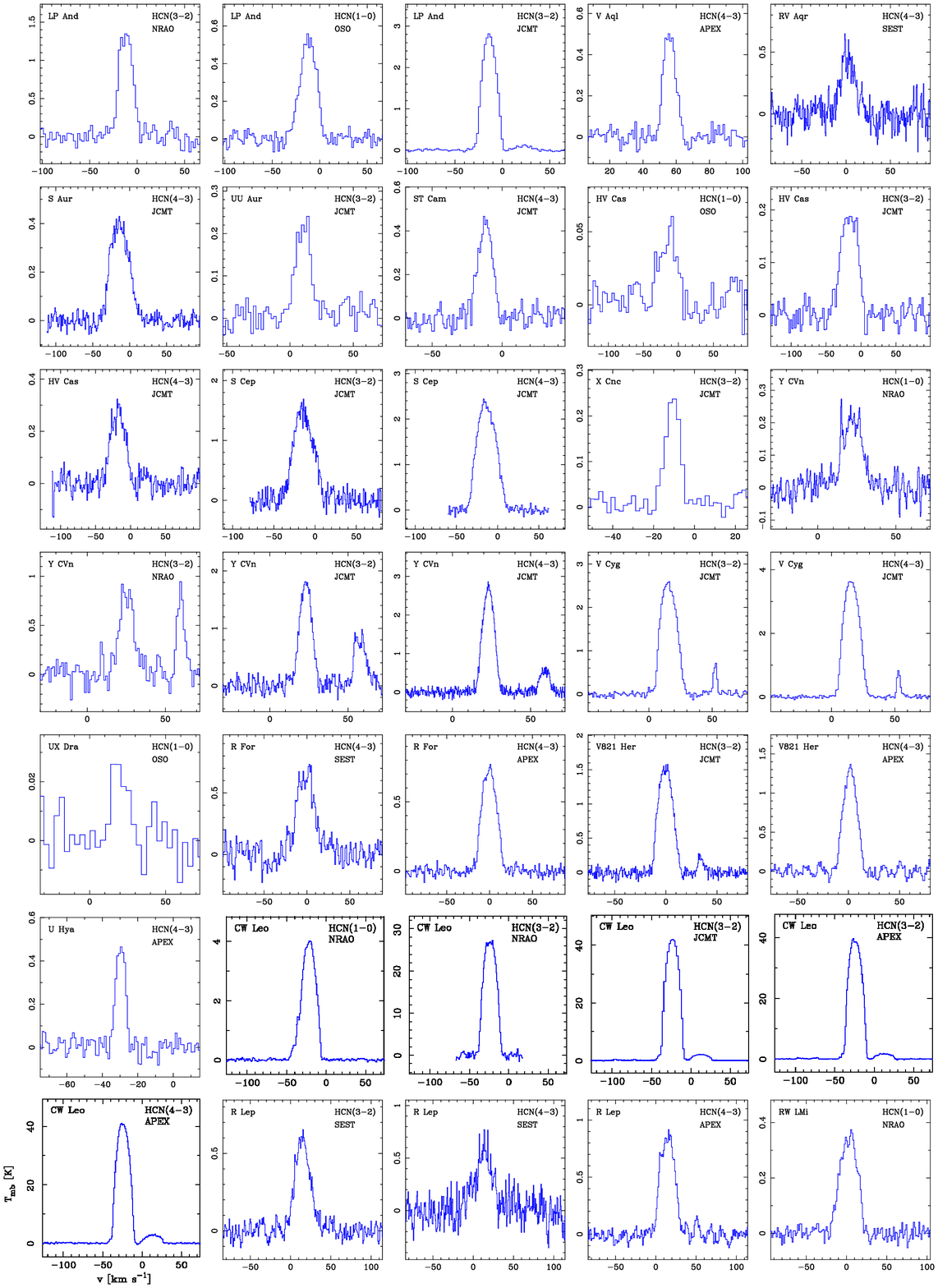}}
  \caption{New multi-transition HCN observations of carbon stars.} 
  \label{spectrac1}
\end{figure*}

\begin{figure*}
\centerline{\includegraphics[width=14cm]{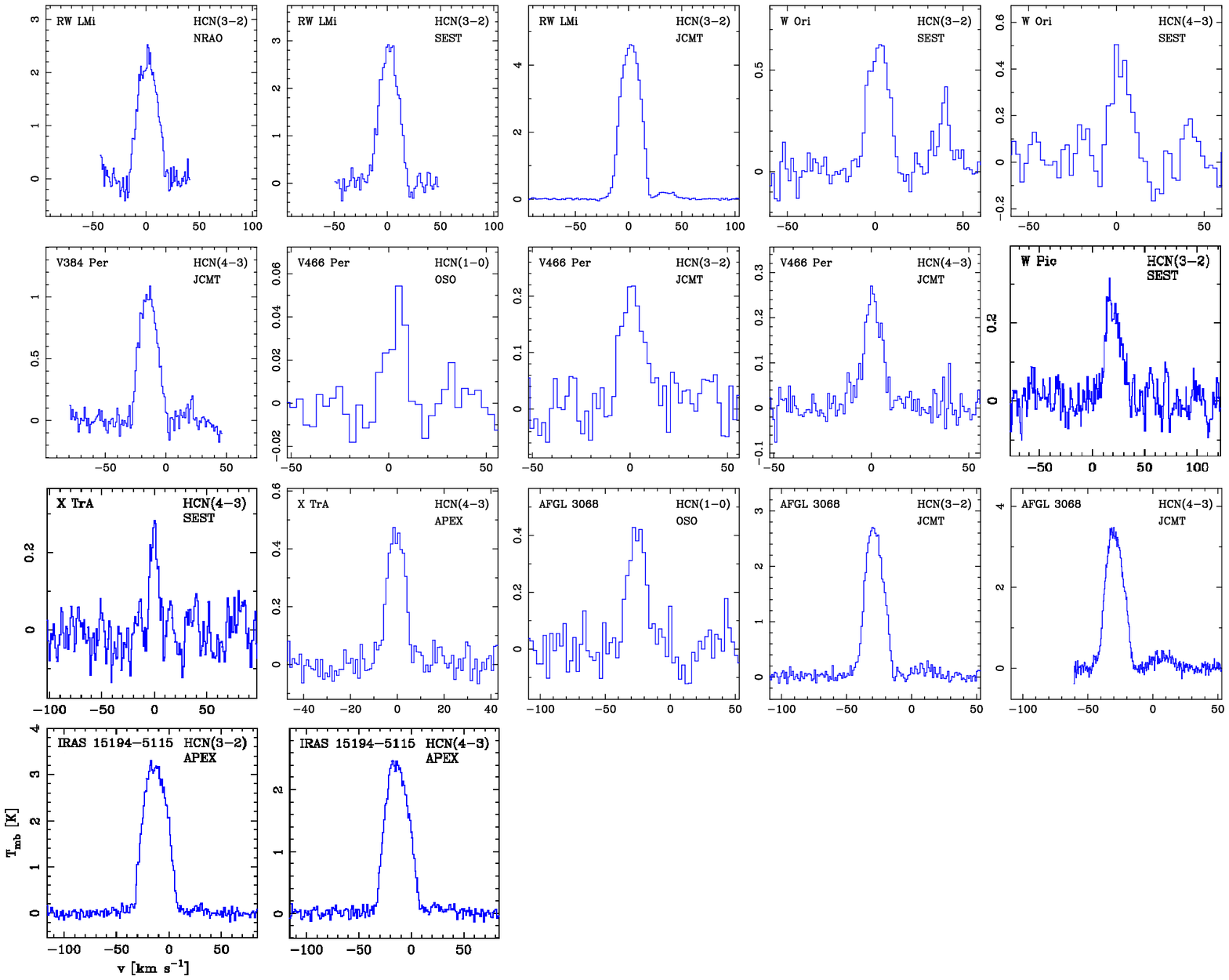}}
  \caption{New multi-transition HCN observations of carbon stars.} 
  \label{spectrac2}
\end{figure*}

\begin{figure*}
\centerline{\includegraphics[width=17cm]{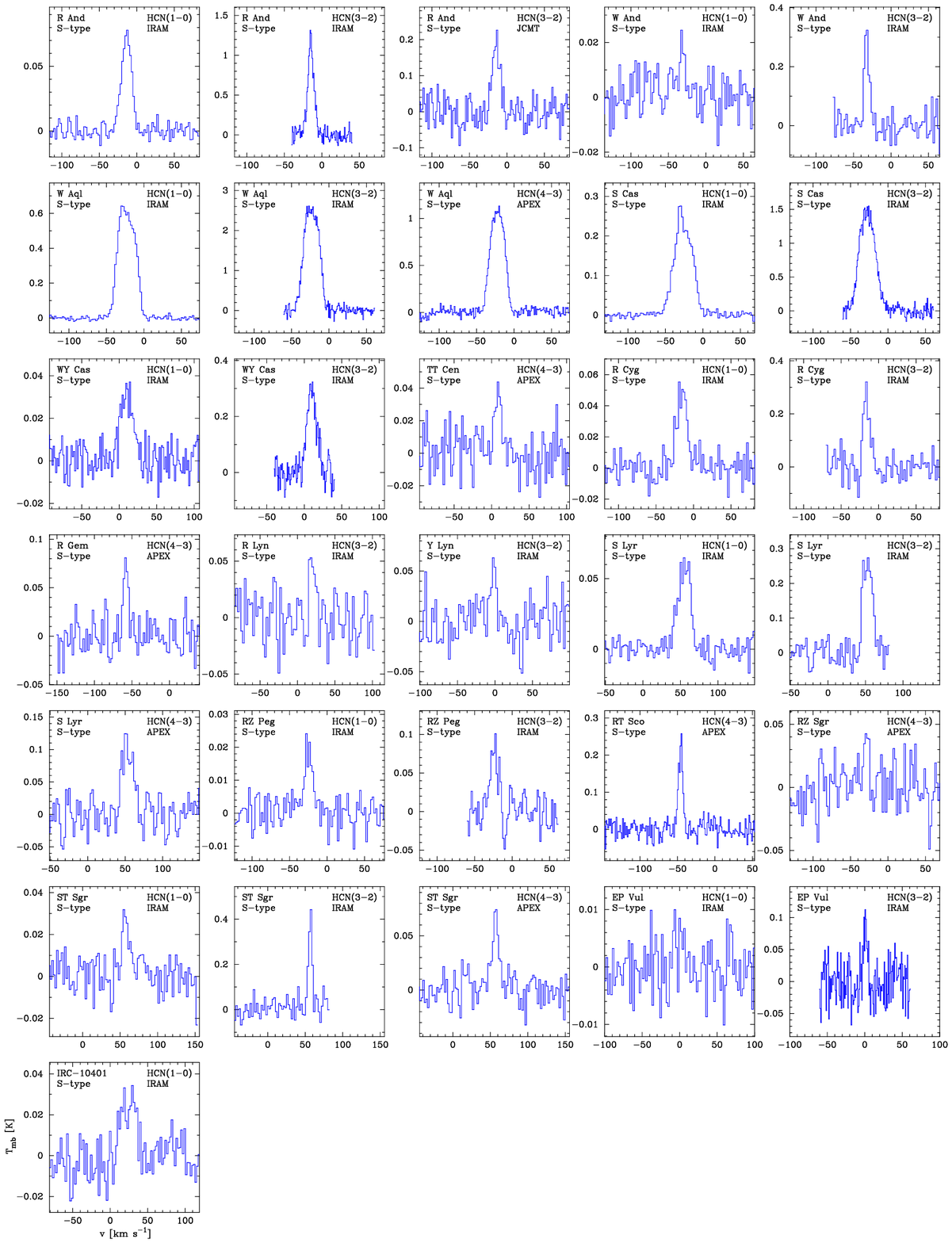}}
  \caption{New multi-transition HCN observations of S-type stars.} 
  \label{spectras}
\end{figure*}

\begin{figure*}
\centerline{\includegraphics[width=14cm]{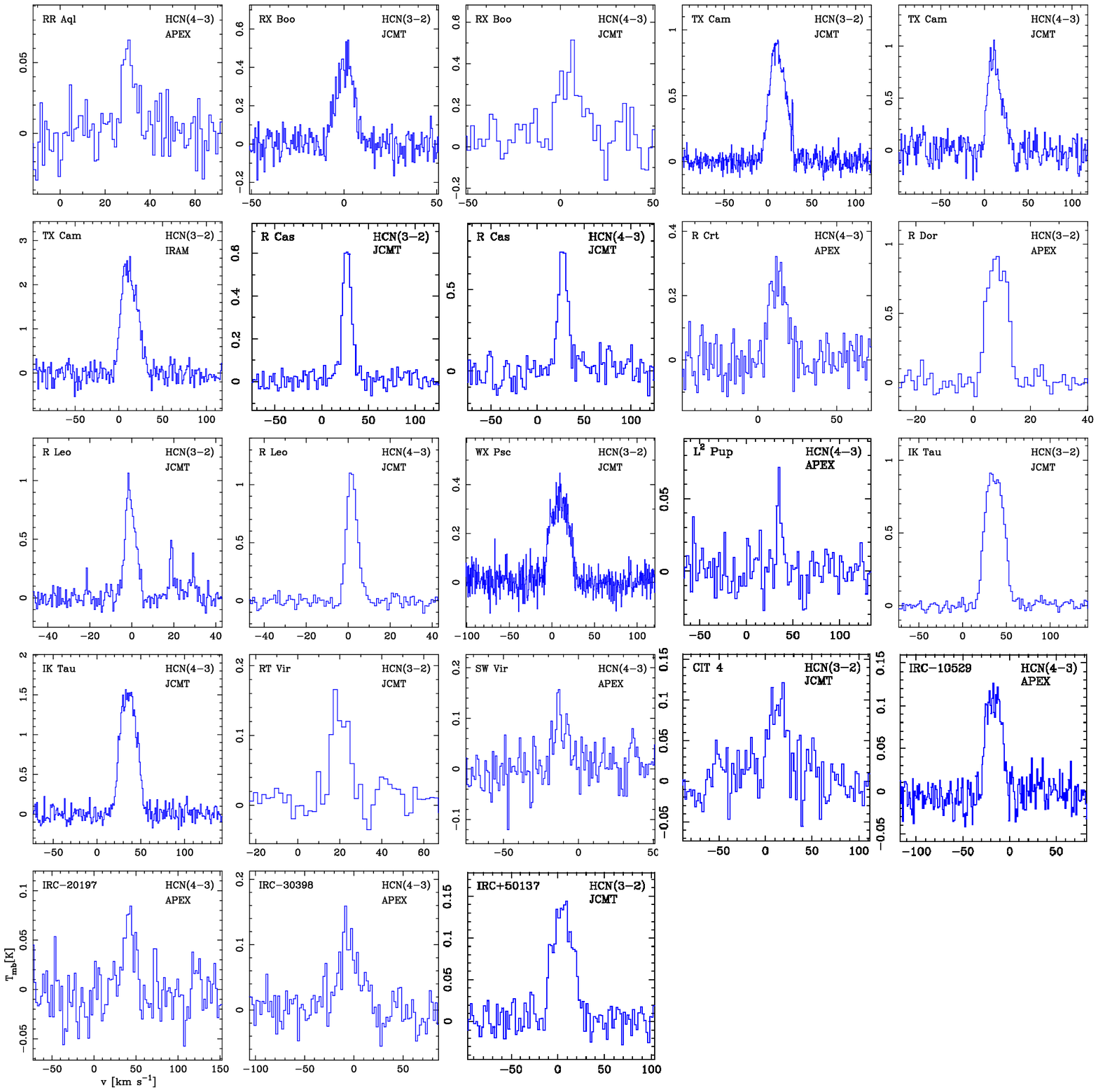}}
  \caption{New multi-transition HCN observations of M-type stars.} 
  \label{spectram}
\end{figure*}

The spectra obtained from our new multi-transition HCN survey are shown in Figs~\ref{spectrac1}--\ref{spectram}. See Sect.~\ref{single-dish} for details on the observations.
The corresponding (velocity) integrated intensities, $\int T_{\mathrm{mb}}\, dv$, are reported in Table~\ref{intensities1} \& \ref{intensities2}, including also previously detected lines that are used in the HCN excitation analysis.


\begin{table*}
\caption{Integrated ($I_{\mathrm{obs}} = \int T_{\mathrm{mb}}\,dv$) line intensities in K\,km\,s$^{-1}$ for the HCN ($v$\,$=$\,0, $J$\,$\rightarrow$\,$J-1$) line emission used in the excitation analysis.}
\label{intensities1}
$
\begin{array}{p{0.17\linewidth}ccccccccccccccccccccc}
\hline
\noalign{\smallskip}
\multicolumn{1}{c}{{\mathrm{Source}}} &
\multicolumn{2}{c}{{\mathrm{NRAO}}} & &
\multicolumn{1}{c}{{\mathrm{OSO}}} & &
\multicolumn{3}{c}{{\mathrm{SEST}}}  &&
\multicolumn{2}{c}{{\mathrm{JCMT}}}  &&
\multicolumn{2}{c}{{\mathrm{HHT}}}  &&
\multicolumn{2}{c}{{\mathrm{APEX}}} &&
\multicolumn{2}{c}{{\mathrm{IRAM}}}  \\ 
\cline{2-3}
\cline{5-5}
\cline{7-9}
\cline{11-12}
\cline{14-15}
\cline{17-18}
\cline{20-21}
&
\multicolumn{1}{c}{1\rightarrow0} & 
\multicolumn{1}{c}{3\rightarrow2} &&
\multicolumn{1}{c}{1\rightarrow0}& &
\multicolumn{1}{c}{1\rightarrow0} &
\multicolumn{1}{c}{3\rightarrow2} &
\multicolumn{1}{c}{4\rightarrow3} &&
\multicolumn{1}{c}{3\rightarrow2} &
\multicolumn{1}{c}{4\rightarrow3} &&
\multicolumn{1}{c}{3\rightarrow2} &
\multicolumn{1}{c}{4\rightarrow3} &&
\multicolumn{1}{c}{3\rightarrow2} &
\multicolumn{1}{c}{4\rightarrow3} &&
\multicolumn{1}{c}{1\rightarrow0} & 
\multicolumn{1}{c}{3\rightarrow2}
\\
\noalign{\smallskip}
\hline
\noalign{\smallskip}
 &&&&&&&&&&&&&&&&&&& \\
{\em Carbon stars} \\
\ \object{LP And} 	& \cdots & \phantom{0}24.1^1 && \phantom{0}11.0^1  && \cdots & \cdots &  \cdots && \phantom{0}51.1^1  &   \cdots &&  \cdots &  \cdots && \cdots & \cdots && \phantom{0}27.7^6 & \cdots  \\
\ \object{V Aql}   	& \cdots & \cdots && \phantom{00}1.5^2 && \phantom{00}1.3^2 & \cdots & \cdots && \cdots & \cdots && \cdots & \cdots && \cdots & \phantom{00}4.6^1 && \cdots & \cdots \\
\ \object{RV Aqr}	& \cdots & \cdots && \cdots && \phantom{00}2.3^2 & \phantom{0}11.4^2 & \phantom{00}9.1^1 && \cdots & \cdots && \cdots & \cdots && \cdots & \cdots && \cdots & \cdots \\
\ \object{S Aur}  	& \cdots & \cdots && \phantom{00}2.9^2 && \cdots & \cdots & \cdots && \cdots & \phantom{0}12.2^4 && \cdots & \cdots && \cdots & \cdots && \cdots & \cdots \\
\ \object{UU Aur} 	& \cdots & \cdots && \phantom{00}5.9^2 && \cdots & \cdots & \cdots && \phantom{00}3.5^1 & \cdots && \cdots & \cdots && \cdots & \cdots && \phantom{0}21.6^6 & \cdots \\
\ \object{ST Cam} 	& \cdots & \cdots && \phantom{00}2.4^2 && \cdots & \cdots & \cdots && \cdots & \phantom{00}5.4^4 && \cdots & \cdots && \cdots & \cdots && \cdots & \cdots \\
\ \object{HV Cas}  	& \cdots & \cdots && \phantom{00}1.4^1 && \cdots & \cdots & \cdots && \phantom{00}5.6^1 & \phantom{00}7.0^1 && \cdots & \cdots  && \cdots &  \cdots && \cdots & \cdots \\
\ \object{S Cep}		& \cdots & \cdots && \phantom{0}12.9^2 && \cdots & \cdots & \cdots && \phantom{0}43.4^4 & \phantom{0}66.4^4 && \phantom{0}17.8^3 &  \phantom{00}30.6^3 && \cdots & \cdots && \phantom{0}35.2^6 & \cdots \\
\ \object{X Cnc}  	& \cdots & \cdots && \phantom{00}1.1^2 && \cdots & \cdots & \cdots && \phantom{00}2.2^1 & \cdots && \cdots & \cdots && \cdots & \cdots && \cdots & \cdots  \\
\ \object{Y CVn}        & \phantom{0}3.1^1 & \phantom{00}9.6^1 && \phantom{00}9.4 ^2 && \cdots & \cdots & \cdots && \phantom{0}17.2^1 & \phantom{0}24.8^1 && \cdots & \cdots && \cdots & \cdots && \phantom{0}25.4^6 & \cdots \\
\ \object{V Cyg}        &\cdots & \cdots && \phantom{00}8.0^2 && \cdots & \cdots & \cdots && \phantom{0}39.8^4 & \phantom{0}56.0^4 && \phantom{0}12.9^3 & \phantom{00}31.9^3 && \cdots  & \cdots && \phantom{0}20.9^6 & \cdots \\
\ \object{UX Dra}  	& \cdots & \cdots && \phantom{00}0.3^1 && \cdots & \cdots & \cdots && \cdots & \cdots && \cdots & \cdots && \cdots & \cdots && \cdots & \cdots \\
\ \object{R For}    	& \cdots & \cdots && \cdots && \phantom{00}2.0^2 & \phantom{0}11.8^2 & \phantom{0}16.5^1 && \cdots & \cdots && \cdots & \cdots && \cdots & \phantom{0}15.5^1 && \cdots & \cdots \\
\ \object{V821 Her}    & \cdots & \cdots && \phantom{00}7.0^1 && \cdots & \cdots & \cdots && \phantom{0}26.6^1 & \cdots && \cdots & \cdots && \cdots & \phantom{0}21.4^1 && \phantom{0}12.8^6 & \cdots \\
\ \object{U Hya}    	& \cdots & \cdots && \cdots && \phantom{00}1.3^2 & \phantom{00}3.5^2 & \cdots && \cdots & \cdots && \cdots & \cdots && \cdots & \phantom{00}3.1^1 && \cdots & \cdots \\
\ \object{CW Leo}  	& 84.0^1 & 547.7^1 && 277.5^2 && 173.3^2 & 595.6^2 & \cdots && 884.2^1 & \cdots && 481.0^3 & 1010.0^3 && 820.4^1 & 833.2^1 && 431.0^6 & \cdots \\   
\ \object{R Lep}     	& \cdots & \cdots && \cdots && \phantom{00}2.3^2 & \phantom{0}13.8^1 & \phantom{0}15.0^1  && \cdots & \cdots && \cdots & \cdots && \cdots & \phantom{0}18.8^1 && \cdots & \cdots \\
\ \object{RW LMi}  	& \phantom{0}9.2^1 & \phantom{0}48.7^1 && \phantom{0}26.4^2 && \phantom{0}22.4^2 & \phantom{0}61.8^1 & \cdots && 106.2^1 & \cdots  && \phantom{0}43.1^3 & \phantom{00}90.0^3 && \cdots & \cdots && \phantom{0}85.4^6 & \cdots  \\   
\ \object{W Ori}		& \cdots & \cdots && \phantom{00}4.6^2 && \phantom{00}4.9^2 & \phantom{00}4.5^1 & \phantom{00}5.9^1  && \cdots & \cdots && \cdots & \cdots && \cdots & \cdots && \cdots & \cdots \\
\ \object{V384 Per} 	& \cdots & \cdots && \phantom{00}6.2^2 && \cdots   & \cdots &  \cdots && \cdots & \phantom{0}18.3^1 &&  \cdots & \cdots &&  \cdots & \cdots && \cdots & \cdots  \\
\ \object{V466 Per}	& \cdots & \cdots && \phantom{00}0.4^1 && \cdots & \cdots & \cdots && \phantom{00}2.4^1 & \phantom{00}3.0^1 && \cdots & \cdots && \cdots & \cdots && \cdots & \cdots \\
\ \object{W Pic}  	& \cdots & \cdots && \cdots && \phantom{00}0.9^2 & \phantom{00}4.6^1 & \cdots && \cdots & \cdots && \cdots & \cdots && \cdots & \cdots && \cdots & \cdots \\
\ \object{X TrA} 	& \cdots & \cdots && \cdots && \phantom{00}3.1^2 & \phantom{00}4.7^2 & \phantom{00}2.3^1 && \cdots & \cdots && \cdots & \cdots && \cdots & \phantom{00}4.4^1 && \cdots & \cdots \\
\ \object{R Vol}     	& \cdots & \cdots && \cdots && \phantom{00}1.9^2 & \phantom{0}12.8^2 & \cdots && \cdots & \cdots && \cdots & \cdots && \cdots & \cdots && \cdots & \cdots \\
\ \object{AFGL\,3068}	& \cdots & \cdots && \phantom{00}7.0^1 && \cdots & \cdots & \cdots && \phantom{0}46.3^1 & \phantom{0}56.9^1 && \cdots & \cdots && \cdots & \cdots && \phantom{0}31.8^6 & \cdots \\
\ \object{IRAS\,15194-5115} & \cdots & \cdots && \cdots && \phantom{0}17.2^5 & 139.8^7 & \cdots && \cdots & \cdots && \cdots & \cdots && \phantom{0}89.0^1 & \phantom{0}66.2^1 && \cdots & \cdots \\
 &&&&&&&&&&&&&&&&&&& \\
{\em S-type stars}\\
\ \object{R And}  	& 0.3^7 & \cdots && \cdots && \cdots & \cdots & \cdots && 2.4^1 & \cdots && \phantom{00}0.9^3 & \phantom{00}2.3^3 && \cdots & \cdots && \phantom{0}1.2^1\phantom{0} & 11.7^1 \\
\ \object{W And} 	& \cdots & \cdots && \cdots && \cdots & \cdots & \cdots && \cdots & \cdots && \phantom{00}0.5^3 & \cdots && \cdots & \cdots && \phantom{0}0.2^1\phantom{0} & \phantom{0}2.5^1 \\
\ \object{W Aql }   	& 2.8^7 & \cdots && \cdots && 4.5^8 & \cdots & \cdots && \cdots & \cdots && \phantom{00}9.9^3 & \phantom{0}11.8^3 && \cdots & \phantom{0}25.5^1 && 18.0^1\phantom{0} & 63.4^1 \\
\ \object{S Cas}       	& 1.4^7 & \cdots && \cdots && \cdots & \cdots & \cdots && \cdots & \cdots && \phantom{00}3.9^3 &\cdots && \cdots & \cdots && \phantom{0}7.2^1\phantom{0} & 38.1^1 \\
\ \object{WY Cas}   	& \cdots & \cdots && \cdots && \cdots & \cdots & \cdots && \cdots & \cdots && \cdots & \cdots && \cdots & \cdots && \phantom{0}0.6^1\phantom{0} & \phantom{0}4.5^1 \\
\ \object{TT Cen}    	& \cdots & \cdots && \cdots && \cdots & \cdots & \cdots && \cdots & \cdots && \cdots & \cdots && \cdots & \phantom{00}0.4^1 && \cdots & \cdots \\
\ \object{R Cyg}       	& \cdots & \cdots && \cdots && \cdots & \cdots & \cdots && \cdots & \cdots && \phantom{00}0.8^3 & \phantom{00}1.2^3 && \cdots & \cdots  && \phantom{0}0.8^1\phantom{0} & \phantom{0}2.9^1 \\
\ \object{$\chi$ Cyg}	& 3.0^7 & \cdots && \cdots && \cdots & \cdots & \cdots && \cdots & \cdots && \phantom{0}12.3^3 & \phantom{0}29.0^3 && \cdots & \cdots && 23.1^6\phantom{0} & \cdots \\
\ \object{R Gem}   	& \cdots & \cdots && \cdots && \cdots & \cdots & \cdots && \cdots & \cdots && \cdots & \cdots && \cdots & \phantom{00}0.6^1 && \cdots & \cdots \\
\ \object{R Lyn}       	& \cdots & \cdots && \cdots && \cdots & \cdots & \cdots && \cdots & \cdots && \cdots & \cdots && \cdots & \cdots && \cdots & \phantom{0}0.4^1 \\
\ \object{Y Lyn}       	& \cdots & \cdots && \cdots && \cdots & \cdots & \cdots && \cdots & \cdots && \cdots & \cdots && \cdots & \cdots && \cdots & \phantom{0}0.5^1 \\
\ \object{S Lyr}      	& \cdots & \cdots && \cdots && \cdots & \cdots & \cdots && \cdots & \cdots && \cdots & \cdots && \cdots & \phantom{00}1.8^1 && \phantom{0}1.2^1\phantom{0} & \phantom{0}4.1^1 \\
\ \object{RZ Peg}    	& \cdots & \cdots && \cdots && \cdots & \cdots & \cdots && \cdots & \cdots && \cdots & \cdots && \cdots & \cdots && \phantom{0}0.3^1\phantom{0} & \phantom{0}1.1^1 \\
\ \object{RT Sco} 	& \cdots & \cdots && \cdots && 0.2^8 & \cdots & \cdots && \cdots & \cdots && \cdots & \cdots && \cdots & \phantom{00}1.6^1 && \cdots & \cdots \\
\ \object{RZ Sgr}    	& \cdots & \cdots && \cdots && \cdots & \cdots & \cdots && \cdots & \cdots && \cdots & \cdots && \cdots & \phantom{00}0.3^1 && \cdots & \cdots \\
\ \object{ST Sgr}    	& \cdots & \cdots && \cdots && \cdots & \cdots & \cdots && \cdots & \cdots && \cdots & \cdots && \cdots & \phantom{00}0.8^1 && \phantom{0}0.3^1\phantom{0} & \phantom{0}2.5^1 \\
\ \object{EP Vul}     	& \cdots & \cdots && \cdots && \cdots & \cdots & \cdots && \cdots & \cdots && \cdots & \cdots && \cdots & \cdots && \phantom{0}0.06^1 & \phantom{0}0.5^1 \\
\ \object{IRC-10401}	& \cdots & \cdots && \cdots && \cdots & \cdots & \cdots && \cdots & \cdots && \cdots & \cdots && \cdots & \cdots && \phantom{0}0.6^1\phantom{0} & \cdots \\
 &&&&&&&&&&&&&&&&&&& \\
\noalign{\smallskip}
\hline
\end{array}
$
$^1$ This paper.\\
$^2$ \citet{Olofsson93b}.\\
$^3$ \citet{Bieging00}. Data from the Arizona Radio Observatory Submillimeter Telescope (ARO SMT), previously known as the Heinrich Hertz Submillimeter Telescope (HHT). \\
$^4$ JCMT public archive.\\
$^5$ \citet{Olofsson98b}.\\
$^6$ \citet{Bujarrabal94}.\\
$^7$ \citet{Woods03}.\\
\end{table*}


\begin{table*}
\caption{Integrated ($I_{\mathrm{obs}} = \int T_{\mathrm{mb}}\,dv$) line intensities in K\,km\,s$^{-1}$ for the HCN ($v$\,$=$\,0, $J$\,$\rightarrow$\,$J-1$) line emission used in the excitation analysis.}
\label{intensities2}
$
\begin{array}{p{0.17\linewidth}ccccccccccccccccccccc}
\hline
\noalign{\smallskip}
\multicolumn{1}{c}{{\mathrm{Source}}} &
\multicolumn{2}{c}{{\mathrm{NRAO}}} & &
\multicolumn{1}{c}{{\mathrm{OSO}}} & &
\multicolumn{3}{c}{{\mathrm{SEST}}}  &&
\multicolumn{2}{c}{{\mathrm{JCMT}}}  &&
\multicolumn{2}{c}{{\mathrm{HHT}}}  &&
\multicolumn{2}{c}{{\mathrm{APEX}}} &&
\multicolumn{2}{c}{{\mathrm{IRAM}}}  \\ 
\cline{2-3}
\cline{5-5}
\cline{7-9}
\cline{11-12}
\cline{14-15}
\cline{17-18}
\cline{20-21}
&
\multicolumn{1}{c}{1\rightarrow0} & 
\multicolumn{1}{c}{3\rightarrow2} &&
\multicolumn{1}{c}{1\rightarrow0}& &
\multicolumn{1}{c}{1\rightarrow0} &
\multicolumn{1}{c}{3\rightarrow2} &
\multicolumn{1}{c}{4\rightarrow3} &&
\multicolumn{1}{c}{3\rightarrow2} &
\multicolumn{1}{c}{4\rightarrow3} &&
\multicolumn{1}{c}{3\rightarrow2} &
\multicolumn{1}{c}{4\rightarrow3} &&
\multicolumn{1}{c}{3\rightarrow2} &
\multicolumn{1}{c}{4\rightarrow3} &&
\multicolumn{1}{c}{1\rightarrow0} & 
\multicolumn{1}{c}{3\rightarrow2}
\\
\noalign{\smallskip}
\hline
\noalign{\smallskip}
 &&&&&&&&&&&&&&&&&&& \\
{\em M-type stars} \\
\ \object{RR Aql}   	& \cdots & \cdots && \cdots && \cdots & \cdots & \cdots && \cdots & \cdots && \cdots & \cdots && \cdots & \phantom{00}0.4^1 && \cdots & \cdots \\
\ \object{RX Boo} 	& \cdots & \cdots && \phantom{00}0.5^5 && \cdots & \cdots & \cdots && \phantom{00}4.9^1 & \phantom{00}6.0^1 && \phantom{00}1.8^3 & \phantom{00}4.4^3 && \cdots & \cdots && \phantom{0}0.8^6\phantom{0} & \cdots\\
\ \object{BX Cam}   	& \cdots & \cdots && \phantom{00}1.3^5 && \cdots & \cdots & \cdots && \cdots & \cdots && \phantom{00}2.3^3 & \phantom{00}3.7^3 && \cdots & \cdots && \cdots & \cdots \\
\ \object{TX Cam}  	& \cdots & \cdots && \phantom{00}4.8^5 && \cdots & \cdots & \cdots && \phantom{0}20.1^1 & \phantom{0}16.8^1 && \phantom{00}7.0^3 & \phantom{00}9.5^3 && \cdots & \cdots && 11.3^6\phantom{0} & \phantom{0}53.0^1 \\
\ \object{R Cas}    	& \cdots & \cdots && \phantom{00}1.3^5 && \cdots & \cdots & \cdots && \phantom{00}7.8^1 & \phantom{00}9.0^1 && \phantom{00}4.5^3 &\phantom{0}10.2^3 && \cdots & \cdots && \phantom{0}2.2^6\phantom{0} & \cdots \\
\ \object{R Crt} 		& \cdots & \cdots && \cdots && \phantom{00}0.5^5 & \cdots & \cdots && \cdots & \cdots && \cdots & \cdots && \cdots & \phantom{00}3.3^1 && \cdots & \cdots \\
\ \object{R Dor}      	& \cdots & \cdots && \cdots && \phantom{00}0.6^5 & \cdots & \cdots && \cdots & \cdots && \cdots & \cdots && \phantom{00}7.7^1 & \phantom{00}8.1^1 && \cdots & \cdots \\
\ \object{R Hor}      	& \cdots & \cdots && \cdots && \cdots & \cdots & \cdots && \cdots & \cdots && \cdots & \cdots &&  \cdots & \phantom{00}0.4^1 && \cdots  & \cdots  \\
\ \object{R Hya}     	& \cdots & \cdots && \cdots && \cdots & \cdots & \cdots && \cdots & \cdots && \cdots & \cdots && \cdots & \phantom{00}1.3^1 && \cdots & \cdots \\
\ \object{W Hya}    	& \cdots & \cdots && \cdots && \phantom{00}0.4^5 & \cdots & \cdots && \cdots & \phantom{0}10.8^1 && \phantom{00}6.8^3 & \phantom{0}10.7^3 && \cdots &  \phantom{00}3.1^1 && \cdots & \cdots \\
\ \object{R Leo}     	& \cdots & \cdots && \phantom{00}0.4^5 && \cdots & \cdots & \cdots && \phantom{00}5.5^1 & \phantom{00}6.9^1 && \phantom{00}1.8^3 & \cdots && \cdots & \cdots && \cdots & \cdots \\
\ \object{AP Lyn} 	& \cdots & \cdots && \phantom{00}1.2^5 && \cdots & \cdots & \cdots && \cdots & \cdots && \phantom{00}1.7^3 & \cdots && \cdots & \cdots && \cdots & \cdots \\
\ \object{GX Mon}   	& \cdots & \cdots && \phantom{00}2.3^5 && \cdots & \cdots & \cdots && \cdots & \cdots && \cdots & \phantom{00}5.1^3 && \cdots & \cdots && \cdots &\cdots \\
\ \object{WX Psc}   	& \cdots & \cdots && \phantom{00}3.7^5 && \cdots & \cdots & \cdots && \phantom{00}9.0^1 & \cdots && \phantom{00}5.0^3 &\cdots && \cdots & \cdots && \cdots & \cdots \\
\ \object{L$^2$ Pup}	& \cdots & \cdots && \cdots && \cdots & \cdots & \cdots && \cdots & \cdots && \cdots & \cdots && \cdots & \phantom{00}1.1^1 && \cdots & \cdots \\
\ \object{IK Tau} 	& \cdots & \cdots && \phantom{00}3.5^5 && \cdots & \cdots & \cdots && \phantom{0}23.6^1 & \phantom{0}35.2^1 &&\cdots & \cdots && \cdots & \cdots && 11.2^6\phantom{0} & \cdots \\
\ \object{RT Vir}    	& \cdots & \cdots && \phantom{00}0.3^5 && \cdots & \cdots & \cdots && \phantom{00}1.6^1 & \cdots && \cdots & \cdots && \cdots & \cdots && \cdots & \cdots \\
\ \object{SW Vir}   	& \cdots & \cdots && \cdots && \cdots & \cdots & \cdots && \cdots & \cdots && \phantom{00}0.8^3 & \phantom{00}1.7^3 && \cdots & \phantom{00}1.3^1 && \cdots & \cdots \\
\ \object{CIT 4}     	& \cdots & \cdots && \phantom{00}0.7^5 && \cdots & \cdots & \cdots && \phantom{00}2.3^1 & \cdots && \cdots & \cdots && \cdots & \cdots && \cdots & \cdots \\
\ \object{IRC-10529}    	& \cdots & \cdots && \phantom{00}1.0^5 && \cdots & \cdots & \cdots && \cdots & \cdots && \phantom{00}1.9^3 & \phantom{00}3.9^3 && \cdots &  \phantom{00}2.2^1 && \phantom{0}2.0^6\phantom{0} & \cdots \\
\ \object{IRC-20197}    	& \cdots & \cdots && \cdots && \cdots & \cdots & \cdots && \cdots & \cdots && \cdots & \cdots && \cdots & \phantom{00}1.1^1 && \cdots & \cdots \\
\ \object{IRC-30398}   	& \cdots & \cdots && \cdots && \phantom{00}0.4^5 & \cdots & \cdots && \cdots & \cdots && \phantom{00}1.3^3 & \phantom{00}2.0^3 && \cdots &  \phantom{00}2.7^1 && \cdots & \cdots \\
\ \object{IRC+10365}   	& \cdots & \cdots && \phantom{00}2.0^5 && \cdots & \cdots & \cdots && \cdots & \cdots && \phantom{00}2.1^3 & \phantom{00}4.2^3 && \cdots &  \cdots && \cdots & \cdots \\
\ \object{IRC+40004}   	& \cdots & \cdots && \cdots && \cdots & \cdots & \cdots && \cdots & \cdots && \phantom{00}1.8^3 & \phantom{00}4.7^3 && \cdots & \cdots && \cdots & \cdots \\
\ \object{IRC+50137}   	& \cdots & \cdots && \phantom{00}0.9^5 && \cdots & \cdots & \cdots && \phantom{00}3.6^1 & \cdots && \cdots & \cdots && \cdots & \cdots && \cdots & \cdots \\
 &&&&&&&&&&&&&&&&&&& \\
\noalign{\smallskip}
\hline
\end{array}
$
$^1$ This paper.\\
$^2$ \citet{Olofsson93b}.\\
$^3$ \citet{Bieging00}. Data from the Arizona Radio Observatory Submillimeter Telescope (ARO SMT), previously known as the Heinrich Hertz Submillimeter Telescope (HHT).\\
$^4$ JCMT public archive.\\
$^5$ \citet{Olofsson98b}.\\
$^6$ \citet{Bujarrabal94}.\\
$^7$ \citet{Bieging94}. \\
$^8$ \citet{Bieging98}. \\
\end{table*}


\clearpage

\end{document}